\documentclass[fleqn,usenatbib]{mnras}

\usepackage{newtxtext,newtxmath}
\usepackage[T1]{fontenc}
\usepackage{ae,aecompl}
\usepackage{CJK}

\usepackage{graphicx}	% Including figure files
\usepackage{amsmath}	% Advanced maths commands
\usepackage{amssymb}	% Extra maths symbols
\usepackage{xspace}
\usepackage[normalem]{ulem}
\usepackage[usenames]{color}

\newcommand{\changaMM}{\texttt{{MANGA}}\xspace}
\newcommand{\changa}{{\texttt{ChaNGa}}\xspace}
\newcommand{\be}{\begin{eqnarray}}
\newcommand{\ee}{\end{eqnarray}}

\newcommand{\grad}{\ensuremath{\boldsymbol{\nabla}}}
\newcommand{\vel}{\ensuremath{\boldsymbol{v}}}

\newcommand{\ddt}[1]{\ensuremath{\frac{\partial #1}{\partial t}}}

\newcommand{\state}{\ensuremath{\boldsymbol{\mathcal{U}}}}
\newcommand{\charge}{\ensuremath{\boldsymbol{U}}}

\newcommand{\flux}{\ensuremath{\boldsymbol{\mathcal{F}}}}
\newcommand{\fluxV}{\ensuremath{\boldsymbol{F}}}
\newcommand{\source}{\ensuremath{\boldsymbol{\mathcal{S}}}}
\newcommand{\sourceV}{\ensuremath{\boldsymbol{S}}}

\newcommand{\weight}{\ensuremath{\mathcal{W}}}
\newcommand{\Ir}{\ensuremath{I}}

\newcommand{\Jr}{\ensuremath{J}}
\newcommand{\nr}{\ensuremath{\hat{\boldsymbol{n}}_r}}
\newcommand{\nri}{\ensuremath{\hat{\boldsymbol{n}}_{r,i}}}

\newcommand{\Er}{\ensuremath{E}_r}
\newcommand{\Hr}{\ensuremath{{\boldsymbol{H}}}}
\newcommand{\Fr}{\ensuremath{{\boldsymbol{F}_r}}}
\newcommand{\Kr}{\ensuremath{{\sf K}}}
\newcommand{\Ptr}{\ensuremath{{\sf P}_r}}

\newcommand{\JrCom}{\ensuremath{\tilde{J}}}
\newcommand{\IrCom}{\ensuremath{\tilde{I}}}

\newcommand{\nuCom}{\ensuremath{\tilde{\nu}}}

\newcommand{\rc}{\ensuremath{\bar{c}}}
\newcommand{\Srlab}{\ensuremath{S_r}}
\newcommand{\Srcom}{\ensuremath{\tilde{S}_r}}
\newcommand{\kapAbsPl}{\ensuremath{\kappa_{\rm P}}}
\newcommand{\kapAbsJ}{\ensuremath{\kappa_{\rm J}}}
\newcommand{\kapSca}{\ensuremath{\kappa_{\rm s}}}
\newcommand{\kapAbsPlj}{\ensuremath{\kappa_{\rm P,j}}}
\newcommand{\kapAbsJj}{\ensuremath{\kappa_{\rm J,j}}}
\newcommand{\kapScaj}{\ensuremath{\kappa_{\rm s,j}}}
\newcommand{\wcom}{\ensuremath{\tilde{\weight}_i}}

\newcommand{\ddtCom}[1]{\ensuremath{\frac{\partial #1}{\partial t}}}

\newcommand{\normal}{\ensuremath{\hat{\boldsymbol{n}}}}

\newcommand{\meshv}{\ensuremath{\boldsymbol{w}}}
\newcommand{\facev}{\ensuremath{\boldsymbol{\tilde{w}}_{ij}}}
\newcommand{\facer}{\ensuremath{\boldsymbol{\tilde{r}}_{ij}}}
\newcommand{\meshr}{\ensuremath{\boldsymbol{r}}}

\title{Time Dependent Radiation Hydrodynamics on a Moving Mesh}

\begin{document}

\begin{CJK*}{UTF8}{gbsn}
\author[Chang, Davis, \& Jiang]{Philip Chang$^{1}$, Shane W. Davis$^2$ and Yan-Fei Jiang(姜燕飞)$^3$\\
$^1$ Department of Physics, University of Wisconsin-Milwaukee, 3135 North Maryland Avenue, Milwaukee, WI 53211, USA; chang65@uwm.edu\\
$^2$ Department of Astronomy, University of Virginia,
Charlottesville, VA 22904, USA; swd8g@virginia.edu\\
$^3$ Center for Computational Astrophysics, Flatiron Institute, 162 Fifth Avenue, New York, NY, 10010, USA; yjiang@flatironinstitute.org}

% These dates will be filled out by the publisher
\date{Accepted XXX. Received YYY; in original form ZZZ}

\label{firstpage}
\pagerange{\pageref{firstpage}--\pageref{lastpage}}
\maketitle

\begin{abstract}
We describe the structure and implementation of a radiation hydrodynamic solver for \changaMM, the moving-mesh hydrodynamics module of the large-scale parallel code, Charm N-body GrAvity solver (\changa).  We solve the equations of time dependent radiative transfer using a reduced speed of light approximation following the algorithm of Jiang et al (2014).  By writing the radiative transfer equations as a generalized conservation equation, we solve the transport part of these equations on an unstructured Voronoi mesh.  We then solve the source part of the radiative transfer equations following Jiang et al (2014) using an implicit solver, and couple this to the hydrodynamic equations.  The use of an implicit solver ensure reliable convergence and preserves the conservation properties of these equations even in situations where the source terms are stiff due to the small coupling timescales between radiation and matter.  We present the results of a limited number of test cases (energy conservation, momentum conservation, dynamic diffusion, linear waves, crossing beams, and multiple shadows) to show convergence with analytic results and numerical stability. We also show that it produces qualitatively the correct results in the presence of multiple sources in the optically thin case.
\end{abstract}


\section{Introduction}

% \shane{notes:
%   \begin{itemize}
%   \item Removed reference to sec:recontruction after eq. 8
%
%   \item  In thermal equilib. section: ``we set up a ??''
%
%   \item  ?? in the shadow test section
%
%   \item ``implicit solver determines equilibration explicitly''  Is this a Phil joke?  I changed explitly to directly both places it appeared to avoid any confusion, but if you want to change it back to explicitly, I am ok with it.
%
%   \item I feel like the shadow test might look better in a longer ``box'' but this is somewhat trivial.
%
%   \end{itemize}
%   }

Smooth particle hydrodynamics (SPH) is based upon the Lagrangian view of the Euler equations where the sampling of a fluid is determined from a finite number of particles, and fluid quantities like density and pressure are determined by computing a smoothing kernel over a number of neighbors.  The Lagrangian nature of SPH allows it to conserve linear and angular momentum, but comes at the expense of comparatively poor resolution of shocks due to its smoothing nature.  On the other hand, grid based methods have superior shock capturing abilities due to the use of Godonov schemes, but suffer from grid effects, e.g., the presences of grid direction can affect the conservation of linear and angular momentum.\\

\citet[][hereafter S10]{2010MNRAS.401..791S} developed an arbitrary Lagrangian-Eulerian or moving-mesh (MM) scheme in an effort to capture the best characteristics of both approaches. S10's scheme relies on a Voronoi tessellation to generate well-defined and unique meshes for an arbitrary distribution of points that deform continuously under the movement of the mesh generating points. Implemented into the code, AREPO, it has been used to study a number of different astrophysical problems including cosmological galaxy formation \citep[see for instance][]{2014MNRAS.444.1518V}, disks, and stellar mergers \citep{2015ApJ...806L...1Z,2016ApJ...816L...9O}.

Aside from AREPO, a number of MM codes have been developed based on S10's scheme.  These include TESS \citep{2011ApJS..197...15D}, FVMHD3D \citep{2012ApJ...758..103G}, ShadowFax \citep{2016A&C....16..109V}, RICH \citep{2015ApJS..216...35Y}, DISCO \citep{2016ApJS..226....2D}, and \changaMM \citep{Chang+17}.
These MM schemes have also been extended to include magnetic fields (\citealt{2011MNRAS.418.1392P,2014MNRAS.442...43M,2016MNRAS.463..477M}; Chang 2018, in preparation), better convergence \citep{2016MNRAS.455.1134P,2015MNRAS.452.3853M}, and new physics, such as cosmic rays \citep{2016MNRAS.462.2603P,2017MNRAS.465.4500P}.  In addition, the general scheme of determining the geometry from an arbitrary collection of points has also led to derivative methods such as GIZMO \citep{2015MNRAS.450...53H}.

SPH and MM schemes are similar in their need to compute the nearest neighbors of a point.  As a result of this similarity, the most prolific MM code, AREPO, is built on top of the SPH code, Gadget \citep{2005MNRAS.364.1105S}. While Gadget is perhaps the most widely used SPH code in astrophysics, a similar SPH code, \changa, has been under heavy development over the last decade \citep{Jetley2008,Jetley2010,2015ComAC...2....1M}.  Betraying its galaxy formation origins, \changa includes standard physics modules that have been ported from it predecessor, Gasoline, including metal line cooling, star formation, turbulent diffusion of metals and
thermal energy, and supernovae feedback \citep{2006MNRAS.373.1074S,2010MNRAS.407.1581S}. The most up-to-date description of the SPH algorithms in Gasoline and \changa is given in \citet{Wadsley2017}

\changa is also unique among astrophysical codes in that it uses the Charm++ language and run-time system \citep{KaleKrishnan96} for parallelization rather than a custom message-passing interface design.  While other Charm++ based codes exists, e.g., ENZO and FVMHD3D \citep{2012ApJ...758..103G}, \changa is by far the most mature.  The use of Charm++ allows \changa to demonstrate strong scaling on a single timestepping problem with $12$ billion particles to 512K cores (with 93\% efficiency) and on a multi-timestepping problem with 52 million particles to 128K cores on Blue Waters \citep{2015ComAC...2....1M}.  In the era of exascale computing, such scalability is increasingly important.

Upon the \changa codebase, we have recently developed \changaMM, a MM hydrodynamic solver for \changa \citep{Chang+17}.  \changaMM is largely based on the S10 scheme, but includes a different approach to generating the Voronoi mesh, the use of conserved variables rather than primitives to compute the face states, and various other improvements that have been proposed since the publication of S10.  More recently, we have begun incorporating additional physics into \changaMM that is geared toward the study of dynamical stellar problems, including various equations of states (EOSs) such as the HELMHOLTZ EOS \citep{2000ApJS..126..501T}, the EOS from the stellar evolution code MESA \citep{2011ApJS..192....3P,2013ApJS..208....4P,2015ApJS..220...15P}, and a nuclear equation of state \citep{2010CQGra..27k4103O,2017PhRvC..96f5802S}.  In addition, we have also recently developed a multistepping scheme \citep{2019MNRAS.486.5809P}.

Radiation also plays a role in dynamical stellar problems like tidal disruption of stars, stellar mergers, or common envelope evolution.  To account for the physics of radiation, we implement a radiation hydrodynamics solver in \changaMM.  Radiation hydrodynamics solvers involve the solution to the radiative transfer (RT) equations and a coupling between the radiation and fluid via the momentum and energy equations.

The general RT equations involves seven variables (3+1 for space and time, one for frequency, and two for direction).  As result, the full solution of radiation transfer has traditionally been viewed as computationally intractable.   Hence, most schemes involve a moment formalism to solve the RT equations, which greatly reduced the number of equations that need to be solved.
A number of different closures for the moment equations are possible, including
the flux-limited diffusion (FLD) \citep{1981ApJ...248..321L,2001ApJS..135...95T,2007ApJ...667..626K,2011ApJS..194...23V} and more recently the M1 method \citep{2007A&A...464..429G,2011A&A...529A..35C,2013ApJS..206...21S,2015MNRAS.449.4380R}. The M1 method has also been recently implemented in the MM code, AREPO \citep{2019MNRAS.485..117K}.

The assumption underlying these closures cannot account for arbitrary complex radiation fields and can produce qualitatively incorrect results.
As these problems are rooted in the basic scheme and not mitigated by resolution, alternative schemes have been proposed.  One such scheme is the variable Eddington
tensor (VET) mehod \citep{2012ApJS..199....9D,2012ApJS..199...14J}.  Here, short characteristics are used to compute the specific intensity at every time step, from which direct quadrature
is used to compute the components of the Eddington tensor that is used to close the radiation
moment equations \citep{2003ApJS..147..197H,2012ApJS..199....9D,2012ApJS..199...14J}.

Short characteristics schemes rely on solving the time-independent RT equations, which requires a global iterative solve.  While such a scheme for a regular mesh has been outlined in \citet{2012ApJS..199....9D}, it is far from clear that an equivalent scheme is possible in an unstructured mesh.  The need for a global solve also makes short characteristics schemes difficult to implement.

Recently, \citet[hereafter JSD14]{2014ApJS..213....7J} described
an alternative scheme where they solve the time-dependent RT equation.  The attractiveness of this scheme is that its speed is independent of the number of sources, gives the correct answer in both the optically thin and thick regimes, and is (fairly) straightforward to implement.  Moreover, it can be written as a conservation equation, which makes it easy to adapt to finite volume methods on unstructured meshes and to MMs in particular. As we mentioned above, such schemes have traditionally been viewed as computationally intractable, but advances in computation have made this approach increasingly viable in frequency averaged (gray or multigroup) limits.

In this work, we extend the algorithm of JSD14 to MM and implement it in \changaMM.  The paper is organized as follows.  We discuss radiation hydrodynamics on a MM in \S~\ref{sec:radhydro}, by summarizing the algorithm of \citet{Chang+17}.  We then discuss radiative transfer in \S~\ref{sec:radiation}, including transport (\S~\ref{sec:transport}), the implicit solution for sources (\S~\ref{sec:implicit}), and implementation of boundary conditions (\S~\ref{sec:bc}) and radiation source (S~\ref{sec:sources}).  We then demonstrate the performance of the method in \S~\ref{sec:test problems} with a limited number of test problems,  including radiation and gas thermalization, radiation and gas momentum transfer, crossing beams, and multiple shadows.  We summarize our conclusions and discuss future improvements in \S~\ref{sec:discussion}.

\section{Radiation Hydrodynamics on a Moving Voronoi Tessellation}\label{sec:radhydro}

\changaMM solves the Euler equations and energy evolution equation\footnote{Following S10, we also include an evolution equation for entropy and switch between the two solutions either based on detections of shocks or explicit user input.  In this work, we exclusively use energy evolution.}, which written in conservative form is:
\be
\ddt{\rho} + \grad\cdot\rho\vel &=& 0 \label{eq:continuity}\\
\ddt{\rho\vel} + \grad\cdot\rho\vel\vel + \grad P &=&-\rho\grad\Phi\label{eq:momentum}\\
\ddt{\rho e} + \grad\cdot\left(\rho e + P\right)\vel &=& -\rho\vel\cdot\grad\Phi\label{eq:energy}
\ee
where $\rho$ is the density, $\vel$ is the velocity, $\Phi$ is the gravitational potential, $e= \epsilon + v^2/2$ is the specific energy, $\epsilon$ is the internal energy, and $P(\rho, \epsilon)$ is the pressure.  Equations (\ref{eq:continuity}) - (\ref{eq:energy}) can be written in a compact form by introducing a state vector $\state=(\rho, \rho\vel, \rho e)$:
\be
\ddt{\state} + \int \grad\cdot\flux dV = \source\label{eq:state}
\ee
where $\flux=(\rho\vel, \rho\vel\vel, (\rho e + P)\vel)$ is the flux function, and $\source = (0, -\rho\grad\Phi, -\rho\vel\cdot\grad\Phi$) is the source function.

S10 showed that equation (\ref{eq:state}) can be solved using a finite volume strategy on a moving unstructured mesh.  Moreover, any equation that can be written in this generic form can be solved on moving unstructured meshes.  For instance, the MHD equations can also be cast in this form \citep{2011MNRAS.418.1392P,2011ApJS..197...15D,2012ApJ...758..103G,2014MNRAS.442...43M,2016MNRAS.463..477M}.  We refer the interested reader to S10 and \citet{Chang+17} for a more detailed discussion of the scheme.  Here, we will only briefly describe the scheme to document the algorithm we have implemented and to highlight the differences between the scheme and that of S10 and highlight more recent development since the publication of \citet{Chang+17}.

For each cell, the integral over the volume of the $i$th cell defines the charge of the $i$th cell, $\charge_i$, to be
\be
\charge_i = \int_i \state dV = \state_i V_i,
\ee
where $V_i$ is the volume of the cell.
As in S10, we then use Gauss' theorem to convert the volume integral over the divergence of the flux in equation (\ref{eq:state}) to a surface integral:
\be
\int_i \grad\cdot\flux dV = \int_i \flux\cdot\normal dA
\ee
We now take advantage of the fact that the volumes are Voronoi cells with a finite number of neighbors to define a integrated flux
\be
\sum_{j \in {\rm neighbors}} \fluxV_{ij} A_{ij} = \int_i \flux\cdot\normal dA,
\ee
where $\fluxV_{ij}$ and $A_{ij}$ are the average flux and area of the common face between cells $i$ and $j$.
The discrete time evolution of the charges in the system is given by:
\be
\charge_i^{n+1} = \charge_i^n + \Delta t \sum_j \hat{\fluxV}_{ij} A_{ij} + \Delta t\sourceV_i, \label{eq:time evolution}
\ee
where $\hat{\fluxV}_{ij}$ is an estimate of the half-timestep flux between the initial, $\charge_i^n$, and final states $\charge^{n+1}_i$, and $\sourceV_i^{(n+1/2)} = \int_i \source dV$ is the time-averaged integrated source function.

The steps that we perform to solve equation (\ref{eq:time evolution}) are as follows:
\begin{enumerate}
   \item Estimate the Courant-limited timestep for each cell.  We refer the reader to \citet{Chang+17} for details.  The timestep can either be an individual timestep in a multistep algorithm \citep{2019MNRAS.486.5809P} or a global time step \citep{Chang+17}.
   \item Estimate the half time step state of the cell in two ways. 
   \begin{enumerate}
      \item Use the methodology of S10 as described in \citep{Chang+17}.  Here, we estimate the gradients of the conserved quantities and use these to estimate the half-time-step conserved quantities via equation (\ref{eq:state}).  
      \item Alternatively, solve the RHS of equation (\ref{eq:time evolution}), but with the replacement of $\Delta t \rightarrow \Delta t/2$.  Solve for the fluxes following the methodology described below, but use piecewise continuous reconstruction, instead of the piecewise linear reconstruction done for the full step.  This method follows the time integration method in Athena++ as describe in \citet{2016ApJS..225...22W} and is called the van Leer method. We used this method for this work.  
   \end{enumerate}  
   \item Drift the mesh generating points by a half-time step and recompute the half-time step tessellation.  This is needed to provide second order convergence in time and follows the same idea as \citet{2011ApJS..197...15D} and \citet{2016MNRAS.455.1134P}.
   \item Use the half-time-step state to compute the half-time-step fluxes (described below) and apply the full step.  Include changes due to the source terms (using the half-step values in the van Leer method).
   \item Update the state of the cell to the full step. 
\end{enumerate}
The inclusion of the van Leer half-step prediction is a development in this paper.  The advantage of this method is that source terms are automatically included at second order and it greatly simplifies the code as the equations are only written once as opposed to the previous method where both the integral and differential forms of the equations must be written.  This is especially important in radiation as the source term must be integrated implicitly (as discussed below).  The van Leer method can be easily adapted to multistepping schemes.  The only change is that the ``half-step'' estimate must be taken from the cell's initial state. However, we will only use global time-step in the remainder of this paper.

To estimate the flux across each face, $\hat{\fluxV}_{ij}$, we use an approximate Riemann solver.
As Riemann solvers for irregular cells in multidimensions do not exist, we follow the prescription of S10. We compute the 1-D fluxes across each face in the rest frame of that face and then collectively apply them per timestep.  The steps involved are:
\begin{enumerate}
 \item Estimate the velocity $\facev$ of the face -- following S10, the face velocities are:
 \be\label{eq:face velocity}
 \facev = \frac{(\meshv_i - \meshv_j)\cdot(\facer - 0.5(\meshr_j+\meshr_i))}{|\meshr_j - \meshr_i|}\frac{\meshr_j - \meshr_i}{|\meshr_j - \meshr_i|} + \bar{\boldsymbol{w}}_{ij},
 \ee
 where $\bar{\boldsymbol{w}}_{ij} = 0.5(\meshv_i + \meshv_j)$ is the average velocity of the two mesh generating points and \facer\ is the face center between cells i and j.
\item Estimate the state vector (in the rest frame of the moving face) at the face center (\facer) between the neighboring $i$ and $j$ cells via linear reconstruction. \label{item:half step}
 \item Boost the state vector from the ``lab'' frame to the rest frame of the face center and rotate the state vector such that the x-axis points along the outward normal of the face, i.e., in the direction from $i$ to $j$.
 \item Estimate the flux using an HLL or HLLC (or HLLD for MHD; \citealt{2005JCoPh.208..315M}) Riemann solver implemented following \citet{toro2009riemann}. Here we found that both Riemann solvers give acceptable performance, though the HLL solver is more diffusive for problems that involve large gradients integrated over long timescale, i.e., hydrostatic balance.  By default, we choose HLLC (or HLLD for MHD).
 \item Boost the solved flux back into the ``lab'' frame.
\end{enumerate}

\subsection{Radiative Transfer}\label{sec:radiation}

We are interested in flows that are slow compared to the speed of light.  Thus, we will adopt the reduced speed of light approximation in this paper.  There are two key steps that are required to include radiation in \changaMM.  First, the radiation field, must be solved. Second the source function for the radiation must be computed using the solution of the radiation field. For the first step, the equation of radiative transfer in the reduced speed of light approximation is:
\be\label{eq:radiative transfer}
\ddt{\Ir} + \rc\nr\cdot\grad\Ir = \Srlab,
\ee
where \Srlab\ is the source function in the lab frame and $\rc$ is the reduced speed of light.  For non-relativistic flows, it is usually sufficient to expand out the terms to $O(\beta)$ \citep{1982JCoPh..46...97M} .  An interested reader is referred to JSD14 for an explicit expression of this source term.  Here we refrain from writing the explicit expression, but note that lab frame radiation source term can be rewritten entirely in terms of variables evaluated in the fluid comoving (COM) frame using Lorentz transformations \citep[see also][]{2017arXiv170902845J}.  Here quantities represent frequency integrated (grey) expressions with $I\equiv \int I_\nu d\nu$, but our treatment easily generalizes to multiple independent frequency groups $I_j \equiv \int_{\nu_j}^{\nu_{j+1}} I_\nu d\nu$.

The zeroth, first, and second moments of the intensity, \Ir, are
\be
\Jr &\equiv & \frac {1}{4\pi}\int \Ir d\Omega, \\
\Hr &\equiv & \frac{1}{4\pi}\int \nr \Ir d\Omega,\\
\Kr &\equiv & \frac {1}{4\pi} \int \nr \nr \Ir d\Omega.
\ee
Since the moments are frequency integrated, they relate to the radiation energy density, radiation flux, and radiation pressure tensor via
\be
\Er = \frac{4\pi}{c} \, \Jr,\\
\Fr = 4\pi \, \Hr, \\
\Ptr = \frac{4\pi}{c} \, \Kr.
\ee
Note that $c$ in these equations is {\it by definition} the actual speed of light even when a reduced speed of light $\rc$ is used in equation~(\ref{eq:radiative transfer}). Taking zeroth and first moments of equation (\ref{eq:radiative transfer}), we arrive at the equations for radiation energy and momentum conservation
\be
   \frac{c}{\rc}\ddt{\Er} + \grad \cdot \Fr & = & \int \Srlab d\Omega, \label{eq:rad energy} \\
   \frac{c}{\rc}\ddt{(c^{-2}\Fr)} +\grad \cdot \Ptr & = & \frac{1}{c}\int \Srlab \nr d\Omega  \label{eq:rad momentum},
\ee
which reduce to standard expressions when $\rc=c$.  The quantities inside the time derivatives correspond to the energy and momentum densities of the radiation field. The source terms on the right hand side of this equation represent the net gain/loss of energy and momentum from the gas and the negative of these terms must be added as source terms on the right hand side of the gas energy and momentum equations.  We note that the reduced speed of light approximation does not conserved total energy and momentum between the gas and radiation, which we elaborate on below.

\subsection{Transport Step}\label{sec:transport}

Equation (\ref{eq:radiative transfer}) can be written in terms of a conservation equation, e.g., equation (\ref{eq:state}), which makes it amendable to be solved on a moving Voronoi mesh.  In particular, the state vector for radiation $\state_{\rm rad}=(\Ir_1, . . . , \Ir_N)$, where $N$ is the total number of intensities in angular and frequency space.  The corresponding flux is  $\flux_{\rm rad}=\rc(\hat{\boldsymbol{n}}_{r,1}\Ir_1, . . . ., \hat{\boldsymbol{n}}_{r,N}\Ir_N)$.  Written this way, radiative transfer can be solved in two parts: transport and sources.

The transport step is solved similarly to the hydrodynamic transport step. In particular, the steps are:
\begin{enumerate}
 \item Estimate the velocity $\facev$ of the face -- same as the hydrodynamic steps and given by equation (\ref{eq:face velocity}).
 \item Estimate the half-timestep state vector (in the rest frame of the moving face) at the face center (\facer) between the neighboring $i$ and $j$ cells via linear reconstruction -- same as the hydro step and uses the same limiter as described in \citet{Chang+17}.
 \item Transform the radiative flux to the moving face frame: $\flux = (\rc\nr - \facev) \Ir$.
 \item Calculate flux using a simple upwind solver.
\end{enumerate}
We note that the transport part of the radiative transfer equation is even easier than the hydrodynamic solution.  In particular, no rotations to the moving face is necessary and the fluxes are simple upwind fluxes that do not require an approximate Riemann solver.

\subsection{Implicit Solver and Hydrodynamic Source Terms}\label{sec:implicit}

Having dealt with the transport part of the radiative transfer equation, we now turn to the source terms.  As discussed above, the source terms are usually expanded out to $O(v/c)$ for non-relativistic flows.  However, if we perform Lorentz boosts to and from the fluid COM frame, the system can be solved in a covariant fashion.  In addition to being accurate to all orders in $v/c$, this procedure obviates the need to evaluate higher order angular moments. Here our COM frame is the center of momentum frame for the fluid element.  This approach is the same as described in \citet{2017arXiv170902845J}. The portion of the transfer equation corresponding to the source term update in the lab frame is given by
\be\label{eq:lab radiative source}
\ddtCom{\Ir} = \Srlab = \rc\rho\left(\kapAbsPl \frac{a T^4}{4\pi} + \kapSca \Jr - (\kapAbsJ+\kapSca)\Ir\right),
\ee
where $\kapAbsPl$ and $\kapAbsJ$ are the Planckian and mean absorption opacity respectively.  We assume isotropic scattering (in the comoving frame) with scattering opacity $\kapSca$.  Note that all quantities in the above expression (including the opacities) are defined in the lab frame.  This means that the opacities are not isotropic in this frame and motivates rewriting the source term in terms of COM frame variables.

The Lorentz transformation of the frequency between the lab and COM frame is given by
\be
\frac{\nu}{\nuCom} \equiv \Gamma = \gamma \left(1-\frac{\vel \cdot \nr}{c}\right),
\ee
where $\gamma$ is the Lorentz factor \citep{1986rpa..book.....R}.  The specific intensities $I_\nu$ and extinction coefficients $\alpha_\nu = \kappa_\nu \rho$ then transform according to
\be
\tilde{I}_{\tilde{\nu}} = \Gamma^3 I_\nu,\\
\tilde{\alpha}_{\tilde \nu} = \Gamma^{-1} \alpha_\nu.
\ee
Since our intensities represent frequency integrated quantities $(I=\int I_\nu d\nu)$, they transform with an additional power of $\Gamma$
\be \label{eq:I lab to com}
\IrCom = \Gamma^4 I.
\ee
We also need to transform the weights associated with each angle to the comoving frame since integrals over angle are replaced with sums over these weights.  Weights $\weight$ correspond to solid angle elements $d\Omega$, which transform as $\nu^{-2}$.  Hence,
\be
\wcom = \mathcal{N}\frac{\weight}{\Gamma^2},
\ee
where $\mathcal{N}$ is a normalization factor than ensures that $\sum_i \wcom =1$ and $\JrCom = \sum_i \wcom \IrCom_{i}$, while accounting for the correct relative weighting of different angles in the comoving frame

We can now rewrite equation (\ref{eq:lab radiative source}) using COM frame variables as
\be\label{eq:com radiative source}
\ddtCom{\IrCom} = \Srcom = \Gamma(\nr) \rc\rho\left(\kapAbsPl \frac{a T^4}{4\pi} + \kapSca \JrCom- (\kapAbsJ+\kapSca)\IrCom\right),
\ee
where our convention is to denote comoving frame quantities with a tilde, with the exception of opacities which from this point onward are always defined in the COM frame where they are isotropic.  Note that this is {\it not} the comoving frame transfer equation.  Since the COM frame is a non-inertial frame it would contain additional terms related to acceleration that are not included here.  Since we have instead rewritten the lab frame equation with COM frame variables, we have instead picked up a factor of $\Gamma(\nr)$, where $\nr$ denotes directions in the lab frame.

Frequency and angle integration of equation (\ref{eq:lab radiative source}) yields an expression for the radiation energy equation, which couples to the gas internal energy equation
\be \label{eq:com rad thermal}
\frac{\rho k_B}{\mu(\gamma - 1) m_p}\ddtCom{T} = -c\rho\left(\kapAbsPl {a T^4} - 4\pi\kapAbsJ \JrCom\right).
\label{eq:tgas}
\ee
The relevant timescale for this set of equations is the thermalization time (between gas and radiation) which can be exceedingly short.  As noted in JSD14, to keep the solver stable over reasonable timesteps, the equations must be solved implicitly.

Generalizing to the case with multiple frequency groups, we can proceed to discretize equation (\ref{eq:com radiative source}) with an implicit form as
\be
\frac{\IrCom^{n+1}_{i,j} - \IrCom^{n}_{i,j}}{\Delta t} = \rc\Gamma_i\rho\left(\kapAbsPlj \frac{a (T^{n+1})^4}{4\pi} + \kapScaj \JrCom^{n+1}_{j}- (\kapAbsJj+\kapScaj)\IrCom^{n+1}_{i,j}\right),
\ee
where integers $n$, $i$, and $j$ denote timestep, angle, and frequency, respectively.  Here, we assume that the coupling between different frequency bins occurs only through the $(T^{(n+1)})^4)$ term. Solving for $\IrCom^{n+1}_{i,j}$ yields
\be \label{eq:com radiative source disc}
\IrCom^{n+1}_{i,j} = \frac{\IrCom^{n}_{i,j}+\Delta t \rc \Gamma_i \rho \left(\kapAbsPlj \frac{a}{4\pi}(T^{n+1})^4 + \kapScaj \JrCom^{n+1}_{j}\right)}{1+\Delta t \rc\Gamma_i \rho (\kapAbsJj+\kapScaj)}.
\ee
To proceed, we sum $\IrCom^{n+1}_{i,j}$ over angle, using solid angle weighting factors $\wcom$ to obtain
$\JrCom^{n+1}_j = \sum_i \wcom \IrCom^{n+1}_{i,j}$ and solve for $\JrCom^{n+1}_j$
\be \label{eq:Jcom}
\JrCom^{n+1}_j = \frac{ \Sigma_{\tilde{I}} + \Sigma_\Gamma \Delta t \rc \kapAbsPlj \frac{a}{4\pi}(T^{n+1})^4}
{1- \Delta t\rc\rho (\kapAbsJj+\kapScaj)\Sigma_\Gamma},
\ee
where we have defined
\be
\Sigma_{\tilde{I}} \equiv \sum_i \frac{\IrCom^{n}_{i,j} \wcom}{1+\Delta t \rc (\kapAbsJj+\kapScaj)\Gamma_i},\\
\Sigma_{\Gamma} \equiv \sum_i \frac{\Gamma_i \wcom}{1+\Delta t \rc (\kapAbsJj+\kapScaj)\Gamma_i}.
\ee

We can discretize equation (\ref{eq:com rad thermal}) as
\be \label{eq:com rad thermal disc}
\frac{\rho k_B}{\mu(\gamma - 1) m_p}\frac{T^{n+1}-T^n}{\Delta t} = \sum_j \weight_j c\rho\left(4\pi\kapAbsJj \JrCom^{n+1}_j-\kapAbsPlj {a (T^{n+1})^4}\right),
\ee
where $\weight_j$ denotes possible weights for different frequency groups.  We can now eliminate $\JrCom^{n+1}_j$ using equation (\ref{eq:Jcom}) and solve directly for $T^{n+1}$. For a temperature independent opacity, this yields a quartic equation for $T^{n+1}$, which we solve analytically.  For a temperature dependent opacity, we use Newton-Raphson iteration to arrive at a solution.  We have tested both methods for temperature independent opacities and found that they yield equivalent results.

Finally, we compute the radiation moments for frequency group $j$ and timestep $n$ via
\be
E^{n}_{r,j} = \frac{4\pi}{c} \sum_i \weight_i \Ir^{n}_{i,j},\label{eq:moments energy}\\
\boldsymbol{F}^{n}_{r,j} = 4\pi \sum_i \weight_i \nri \Ir^{n}_{i,j}.\label{eq:moments flux}
\ee

The implementation of the algorithm proceeds as follows:
\begin{enumerate}
\item Compute lab frame values for the radiation energy density $E^{n}_{r,j}$ and flux $\boldsymbol{F}^{n}_{r,j}$ at
  beginning of the source term step using equations (\ref{eq:moments energy}) and (\ref{eq:moments flux}).

\item Transform $\Ir^n_{i,j}$ to $\IrCom^n_{i,j}$ using equation (\ref{eq:I lab to com}), perform the sums over angle needed to evaluate equation (\ref{eq:Jcom}), and solve equation (\ref{eq:com rad thermal disc}) for $T^{n+1}$.

\item For each angle and frequency (if using multiple groups) evaluate $\IrCom^{n+1}_{i,j}$ with equation (\ref{eq:com radiative source disc}) and use equation (\ref{eq:I lab to com}) to transform $\IrCom^{n+1}_{i,j}$ to $\Ir^{n+1}_{i,j}$.

\item Evaluate new lab frame values $E^{n+1}_{r,j}$ and $\boldsymbol{F}^{n+1}_{r,j}$ and compute the changes for the gas energy and momentum via
\be
\Delta E & = & \frac{c}{\rc} \sum_j \weight_j\left( E^{n}_{r,j} - E^{n+1}_{r,j}\right)\label{eq:gas1}\\
\Delta (\rho \vel) & = & \frac{1}{c\rc} \sum_j \weight_j \left(\boldsymbol{F}^{n}_{r,j}-\boldsymbol{F}^{n+1}_{r,j}\right).\label{eq:gas2}
\ee
In other words, any loss of energy or momentum by the radiation field due to the source term coupling must correspond to a gain by gas to ensure energy conservation.  The factors of $c/\rc$ and $1/(c\rc)$ are dictated by the form of equations (\ref{eq:rad energy}) and (\ref{eq:rad momentum}). Finally, we should note that our use of equations (\ref{eq:gas1}) and (\ref{eq:gas2}) conserves $E_g + cE_{\rm rad}/\rc$. Modified schemes can also be used to conserved $E_g + E_{\rm rad}$, but they not as applicable to very optically-thick media.

\end{enumerate}

\subsubsection{Reduced Speed of Light Approximation}\label{sec:reducedc}

The equations above implement the reduced speed of light approximation (RSLA).  In the limit $\rc \rightarrow c$, we recover the standard radiation transfer equations.  Notice that this does not mean the speed of light is reduced in all contexts, as $c$ and $\rc$ both appear in equations in sections (\ref{sec:radiation})-(\ref{sec:implicit}). For example, $c$ appears on the right hand side of equation~(\ref{eq:tgas}) whereas $\rc$ appears on the right hand side of equation (\ref{eq:com radiative source}). The logic of the RSLA is that for some problems the time variability of the radiation field on timescales short compared to the flow time can be incorrect as long as the time variability on driven by the flow is accurate captured.  

This has important implications for energy and momentum conservation.  If we add the radiation source term to the right hand side equation~(\ref{eq:energy}), then equation (\ref{eq:rad energy}) implies
\be
\frac{\partial}{\partial t}\left(\frac{c}{\rc}\Er+\rho e\right) + \grad\cdot\left[\Fr+\left(\rho e + P\right)\vel\right] = -\rho\vel\cdot\grad\Phi.
\ee
Similarly, combining equations~(\ref{eq:momentum}) and (\ref{eq:rad momentum}) yields
\be
\frac{\partial}{\partial t}\left(\frac{\Fr}{c\rc}+\rho\vel\right) + \grad\cdot \left(\Ptr+\rho\vel\vel\right) + \grad P = -\rho\grad\Phi.
\ee
This explains why equations~(\ref{eq:gas1}) and (\ref{eq:gas2}) differ by a factor of $c/\rc$ from standard relations. We only get the standard relations for conservation of energy and momentum in the limit that $\rc \rightarrow c$.  This is the price we pay for ensuring that the quasisteady-state behavior of the radiation on flow timescales is correct.

As discussed in \citet{2013ApJS..206...21S} one must be careful to maintain the appropriate hierarchy of timescales when applying the RSLA.  We can define a characteristic flow timescale of $t_{\rm fl} \simeq L/v$ where $L$ and $v$ are characteristic flow lengths and velocities.  We can also define timescales corresponding to radiation streaming and diffusion as $t_{\rm str} \simeq L/\rc$ and $t_{\rm dif} \simeq L/(\rc \tau)$ where $\tau \ge 1$ is the characteristic optical depth corresponding to length $L$.  The RSLA is unlikely to provide reliable results when $t_{\rm fl} < t_{\rm str}$ so we must generally choose $\rc > v$, where $v$ is the largest flow velocity or sound speed.  In problems with large optical depths, the more constraining relation is that we must also maintain $t_{\rm fl} > t_{\rm dif}$, which requires choosing $\rc > v\tau$.  This generally limits the advantage of applying the RSLA to optically thick flows.

\subsection{Boundary Conditions}\label{sec:bc}

The original implementation of \changaMM \citep{Chang+17} only included periodic boundary conditions.  By comparison AREPO has both reflecting and periodic boundary conditions (S10) and the reflecting boundary conditions can be arbitrarily placed and variable with time.  For the radiation problems, it is useful to have some sort of outflow or radiation boundary condition for the test problems below.  Here, we describe our implementation of outflow or absorbing and radiation boundary conditions as a source implementation.

To implement a form of absorbing boundaries, we mark certain cells as having boundaries where we desire the hydrodynamic and radiative primitive variables to be
\be
\rho\rightarrow \rho_0\quad\textrm{and}\quad \vel \rightarrow 0 \quad\textrm{and}\quad \Ir \rightarrow I_0 \equiv  \frac{a T_{\rm eff}^4}{4\pi},
\ee
where $I_0$ is the defined boundary radiation specific intensity, which is direction dependent, and $T_{\rm eff}$ is the effective temperature of the radiation source.
To achieve this, we impose the following source terms for the boundary cells:
\be
\ddt{\rho} &=& S_{\rho} = \frac{\rho_0-\rho}{\tau}, \label{eq:rho bound}\\
\ddt{\rho\vel} &=& S_{\rho\vel} = -\frac{\rho\vel}{\tau},\\
\ddt{\Ir} &=& S_{\Ir} = -\frac{\Ir - I_0}{\tau}\label{eq:rad bound},
\ee
where $\tau$ is a relaxation timescale, and is selected to be longer than the timestep, but much smaller than the typical timescales of interest.  We note that the density and velocity source terms produces a source term for the energy as well, which is easily computed.

The discretized version of equations (\ref{eq:rho bound}) - (\ref{eq:rad bound}) gives the timestepping algorithm used in the boundary computation:
\be
\rho^{n+1} &=& \rho^n + \min(\Delta t/\tau, 1)(\rho_0 - \rho^n), \\
\rho^{n+1}\vel^{n+1} &=& \rho^n\vel^n(1 -  \min(\Delta t/\tau, 1))\\
\Ir^{n+1} &=& \Ir^{n} -  (\Ir^{n} - I_0)\min(\Delta t/\tau, 1)),
\ee
where the $\min(\Delta t/\tau, 1)$ limits the correction down to zero per timestep.

\subsection{Radiation Sources}\label{sec:sources}

In addition to the boundary conditions implemented above, we also include radiation source terms in \changaMM. These radiation source terms are defined in the comoving frame and is defined by a variant of equation (\ref{eq:com radiative source}).
\be
\ddtCom{\IrCom} = \Gamma(\nr') \rc\rho\left(\kapAbsPl \frac{a T_{\rm eff}^4}{4\pi} - \kapAbsPl\IrCom\right),
\ee
where $\nr'$ is the user defined directions of interest.  Here the power of the sources are defined in terms of local density and hence require some nontrivial opacity for it to radiate effectively.  The combination of radiation source and radiation boundary conditions (described in \S~\ref{sec:bc}) allow us to study a wide range of different situations.

\section{Test Problems}\label{sec:test problems}

We now demonstrate the accuracy of the radiation hydrodynamics solver in \changaMM with a limited number of test problems.  The implementation of \changaMM currently limits it to 3-D problems.  While one and two dimensional problems are possible for \changaMM in principle, no such extension is planned.
%Non-periodic boundary conditions are also possible, though no such implementation is planned for the near future, again due to lack of demand.  Hence the test problems below are all periodic and 3-D.

\subsection{Thermal Equilibrium}\label{sec:thermal}

The thermal equilibration timescale in many astrophysical contexts can be much shorter than the dynamical time.  Our use of an implicit integrator in the radiation source terms ensures that the solution will be stable for large timesteps (compared to the local thermal time).  As a demonstration of the accuracy and stability of our implicit solver, we setup a 1 pc simulation cube and fill it with $3\times 10^4$ particles.  The box consists of atomic hydrogen with a density of $\rho = 10^{-15}\,{\rm g\,cm}^{-3}$ and initial temperature of $T=10$ K.  The box is also filled with radiation at a initial radiation temperature of $T=200$ K, which is substantially larger than the gas temperature.  We set $\rc = c$. We consider two cases.  In the first case, we consider an equilibration time that is longer than the timestep.  In the second case, we consider an equilibration time much shorter than the timestep. These two cases are achieved for exactly the same initial conditions by changing the opacity of the material, e.g., $\kappa = 1\,{\rm cm^2\,g}^{-1}$ for the first case, $\kappa=100\,{\rm cm^2\,g}^{-1}$ for the second.

\begin{figure}
 \includegraphics[width=0.48\textwidth]{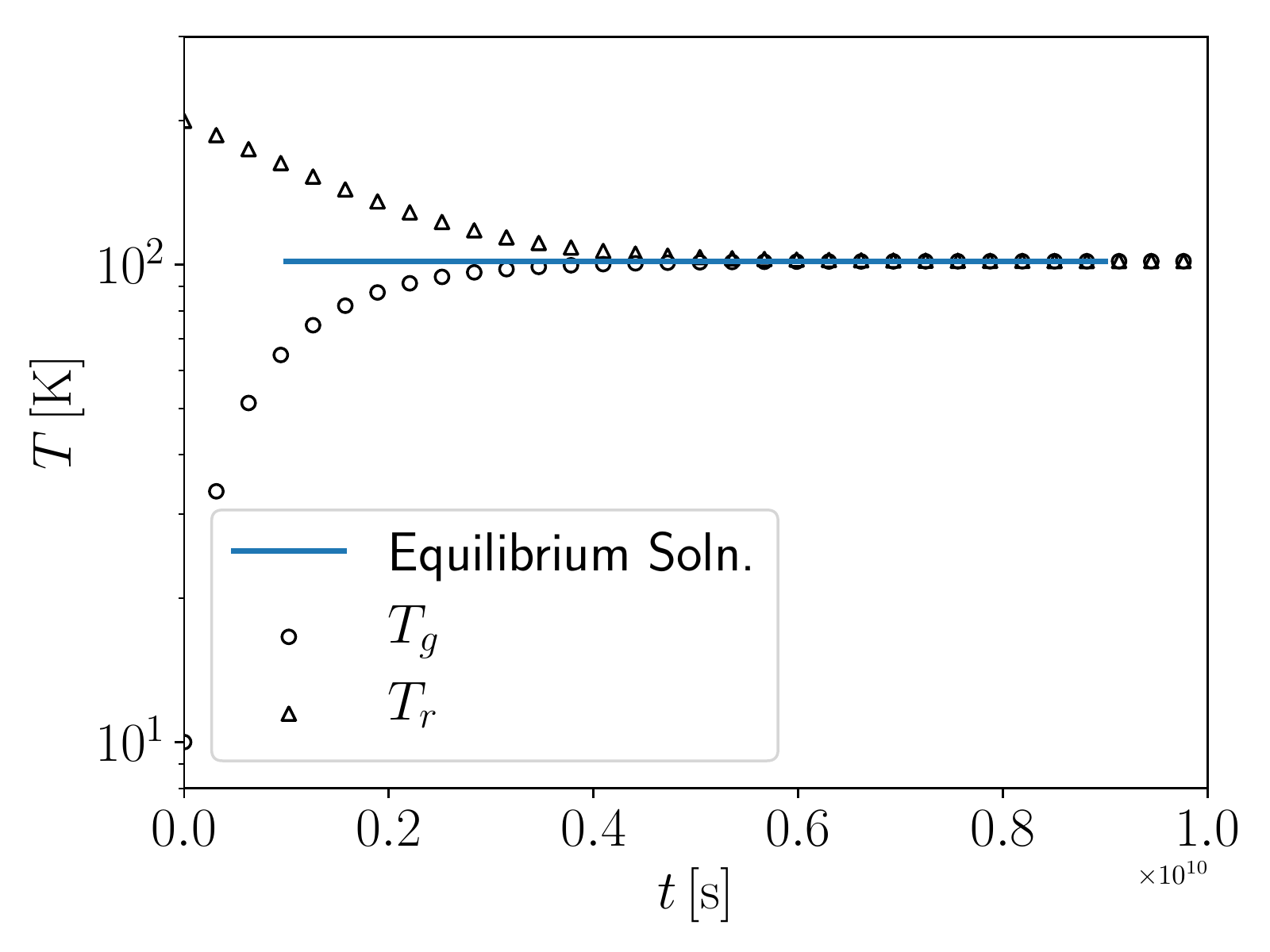}\\
 \includegraphics[width=0.48\textwidth]{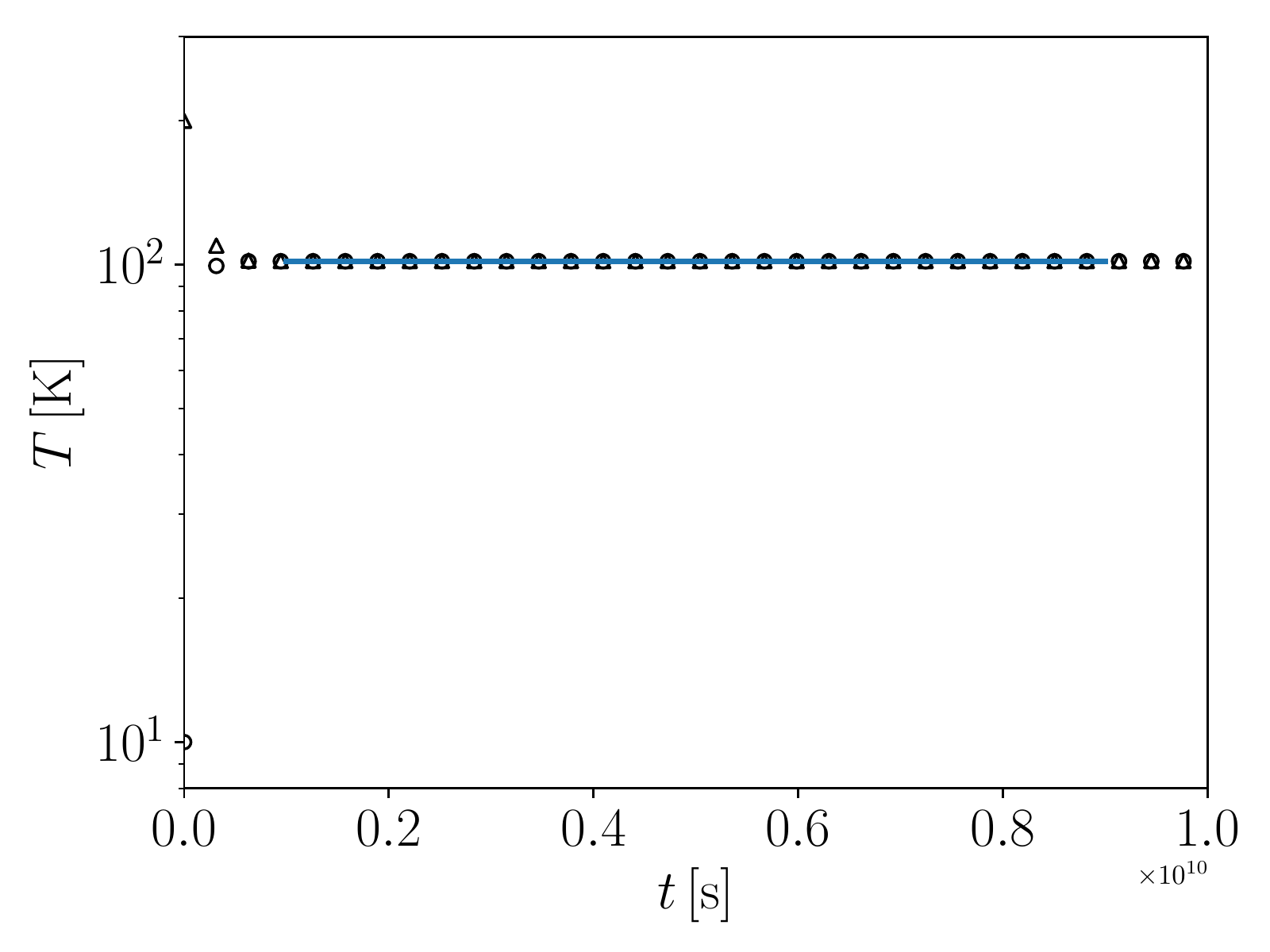}
   \caption{Gas (circles) and Radiation (triangles) Temperatures as a function of t for $\kappa=1\,{\rm cm^2\,g}^{-1}$ (top) and $\kappa=100\,{\rm cm^2\,g}^{-1}$ (bottom).  The initial $T_r = 200$ K, while the initial gas temperature is $T_g = 10$ K.  The solid blue line is the equilibration temperature of $101.7$ K.  \label{fig:thermal}}
\end{figure}

We plot in
Figure \ref{fig:thermal} the evolution of the gas (circles) and radiation (triangles) temperatures as a function of time for $\kappa=1\,{\rm cm^2\,g}^{-1}$ (top plot) and $\kappa=100\,{\rm cm^2\,g}^{-1}$ (bottom plot).  We also plot the expected equilibrium temperature as a solid blue line.  For the $\kappa=100\,{\rm cm^2\,g}^{-1}$ case, the equilibration time is much smaller than the time step, but the implicit solver quickly produces the correct equilibration temperature nevertheless.
This is perhaps unsurprising as the implicit solver determines the equilibration directly.  However, it is reassuring that even in the presence of short thermalization times, \changaMM produces the correct solution.

\subsection{Momentum Conservation}\label{sec:momentum}

\begin{figure}
 \includegraphics[width=0.48\textwidth]{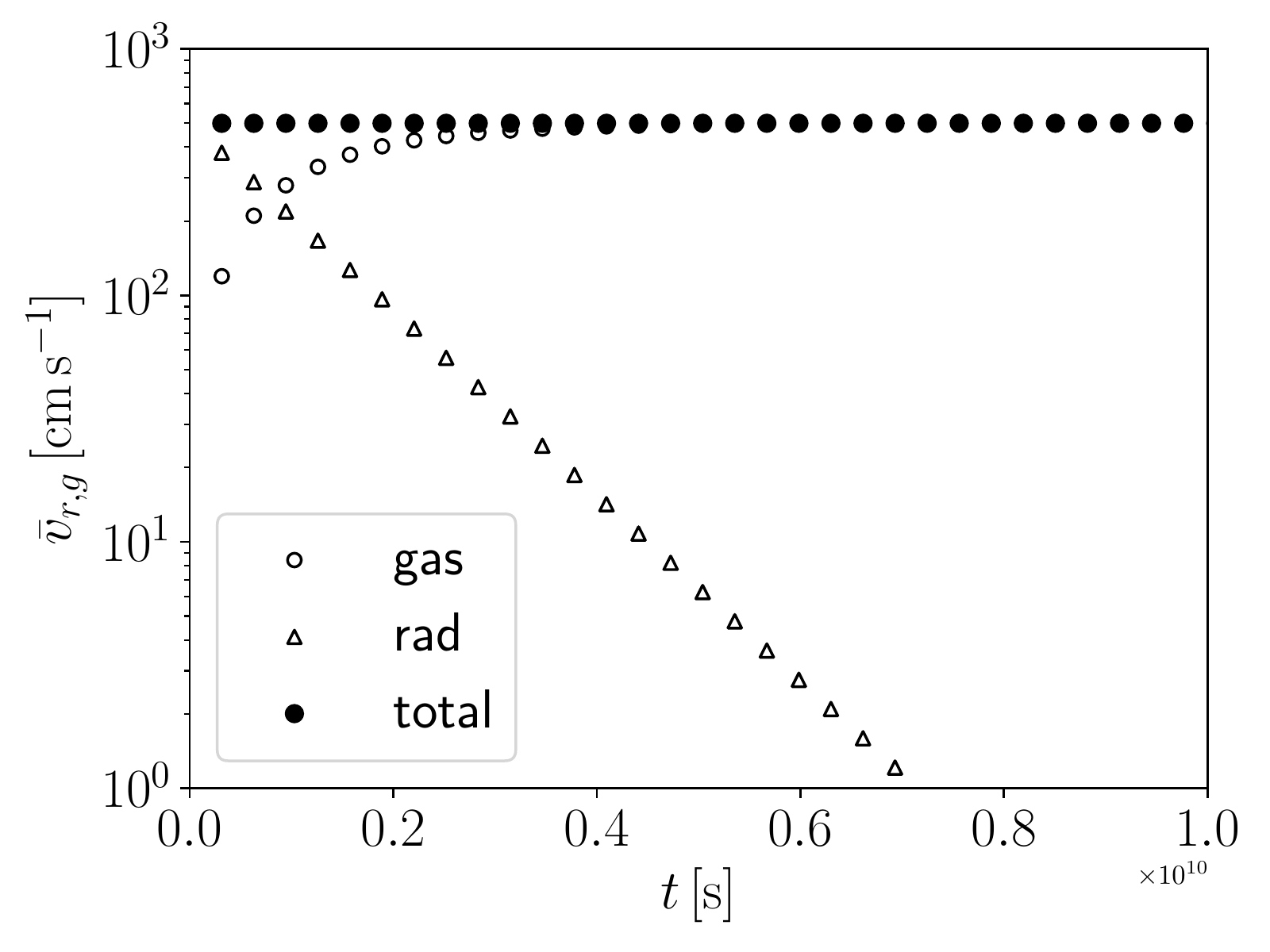}\\
 \includegraphics[width=0.48\textwidth]{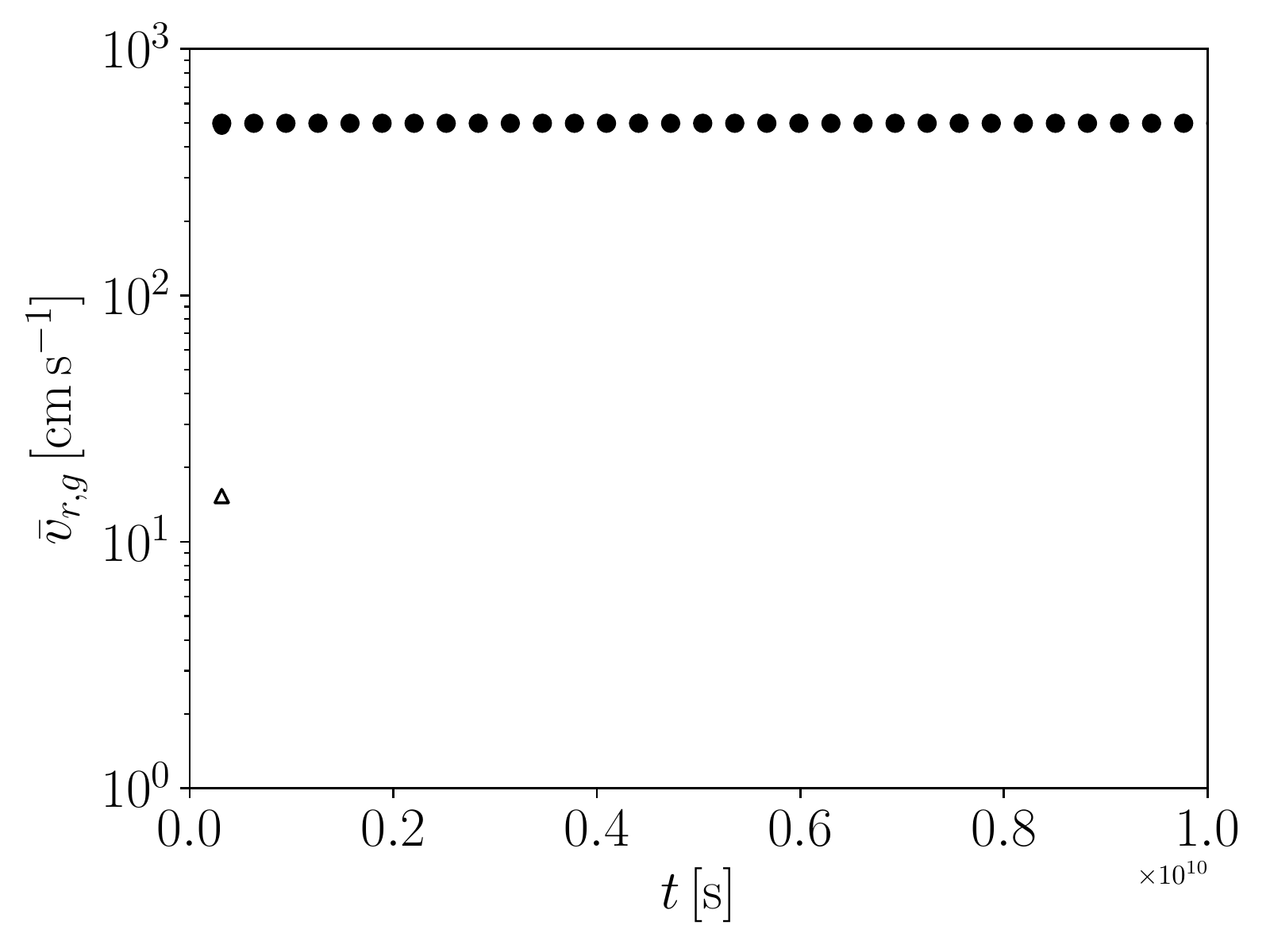}
    \caption{Gas (empty circles) and Radiation (triangles) effective velocities as a function of t for $\kappa=1\,{\rm cm^2\,g}^{-1}$ (top) and $\kappa=100\,{\rm cm^2\,g}^{-1}$ (bottom).  The effective velocity for the total momentum is shown by filled circles. The momentum is conserved to better than $4\times 10^{-4}$.\label{fig:mom}}
\end{figure}

Our scheme also displays excellent momentum conservation properties between radiation and gas.  Similarly to the thermal equilibrium properties discussed above, our use of an implicit integrator in the radiation source terms ensure stability for large timesteps (compared to the local equilibriation timescale).  As an example, we now consider the same case as in \S~\ref{sec:thermal}, but here we impose a radiation flux as oppose to perfectly isotropic radiation.  We do this by imposing that only rays whose normal direction has a $\mu = \cos\theta \ge 0.8$ with the x-axis is populated.   As a result the radiation contains a net momentum, which it will impart on the gas.  Here, we again set $\rc = c$. Again, we consider two cases, the case of long equilibration times, $\kappa = 1\,{\rm cm^2\,g}^{-1}$, and that of short equilibration times, $\kappa = 100\,{\rm cm^2\,g}^{-1}$.  We plot in
Figure \ref{fig:mom} the evolution of the gas (empty circles) and radiation (triangles) effective velocities for radiation, $\bar{v}_r$, and gas, $\bar{v}_g$. as a function of time for $\kappa=1\,{\rm cm^2\,g}^{-1}$ and $\kappa=100\,{\rm cm^2\,g}^{-1}$, where we define
\be
\bar{v}_{\rm r} \equiv m_{\rm tot}^{-1}\int \frac{\Fr}{c^2} dV \quad\textrm{and}\quad \bar{v}_{\rm g} \equiv m_{\rm tot}^{-1}\int {\rho\vel}dV
\ee
We also plot the total momentum as solid black circles which is conserved with a maximum error fraction of $4\times 10^{-4}$.  For the $\kappa=100\,{\rm cm^2\,g}^{-1}$ case, the equilibration time is much smaller than the time step, but the implicit solver produces the proper momentum in the gas.
%Again, we should not be surprised as the implicit solver determines equilibration directly.  However, it is reassuring that these conserved quantities are preserved.

\subsection{Dynamic Diffusion}\label{sec:diff}

\begin{figure}
   \includegraphics[width=0.48\textwidth]{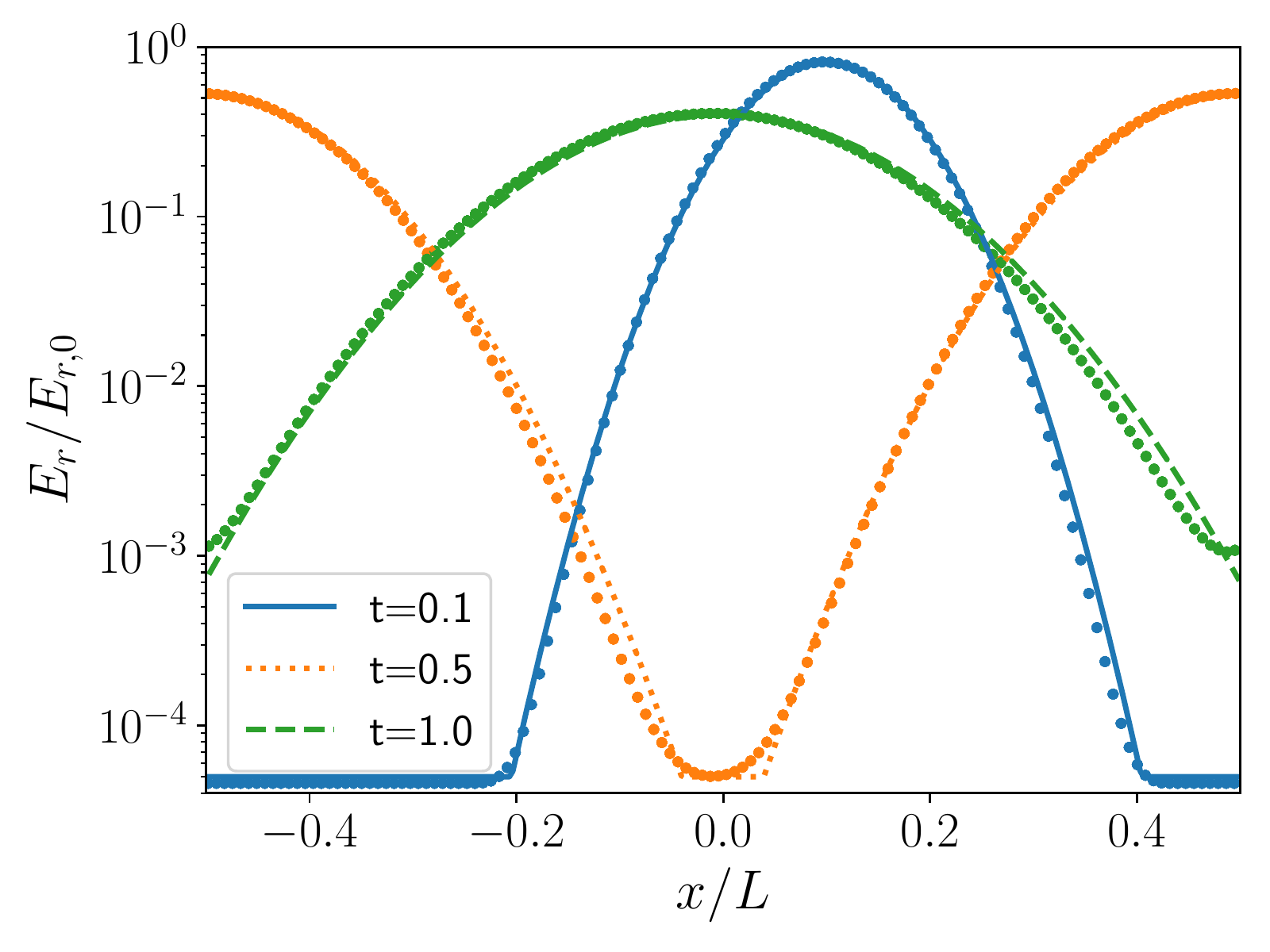}
      \caption{The radiation energy density, $E_r$, as a function of position for three different times of $t = 0.1, 0.5, $ and $1\, L/v_{x,0}$.  The corresponding analytic solutions given by equation (\ref{eq:diff analytic}) are overplotted as solid, dotted, and dashed lines (respectively).\label{fig:diff}}
  \end{figure}

A dynamic diffusion test is useful for testing the diffusion of radiation in an optically thick moving medium where both the advection and diffusion of radiation occurs simultaneously.  Previously, JSD14 argued that the transport part of equation (\ref{eq:radiative transfer}) should be split into a fluid advection term and a radiation transport term to reduce numerical diffusion for optically thick cells. Instead, we have adopted a van Leer second order time integration that allows us to pass this test without using the JSD14 split transport scheme. \footnote{We note that Athena++ also implements a van Leer scheme \citep{2016ApJS..225...22W} and its utility in reducing numerical diffusion in optically thick media has already been demonstrated in that code.

We setup a uniform medium with background density $\rho = 10^{-2}\,{\rm g\,cm}^{-3}$ and box size in the x-direction $L_x = 3\times 10^{10}$ cm and resolve it with 128 cells.  We set $x=0$ to be the center of this domain. The box size in the y and z direction are equal, $L_y=L_z$ and set to be boxes that are $3dx$ across, where $dx$ is the separation of grid points in the x-direction. These physical units are irrelevant to the problem once everything is scaled to the four parameters described below in \S~\ref{sec:linear}, e.g., $\mathcal{C} = c/c_s$, the ratio between the speed of light and isothermal sound speed, $\mathcal{P} = P_{\rm rad}/P_{\rm g}$, the ratio between the radiation and gas pressures, $\tau = \kappa\rho L$, the optical depth in the x-direction.  We setup the mesh-generation points with a regular lattice structure, but perturbed randomly by $10^{-6}$ of the grid spacing to ensure that there are no degeneracies in the Voronoi planes.  The optical depth of the entire box is set to be $\tau = 400$. We set $\rc = c$ and $c_s = 0.1 c$ and the initial velocity of the flow to be $v_{x,0} = 0.1 c$.  We do not evolve the hydrodynamic quantities and allow the mesh to follow the flow.  

The initial profile of the radiation energy density and flux is set to be 
\be
E_r &=& E_{r,0} \exp(-\frac{x^2}{\Delta x^2}),\\
F_r &=& \frac {4v}{c} E_r,
\ee
As in JSD14, we limit the lower value of $E_r$ to be fixed by $E_{r,0}\exp(-10)$ and like the linear wave test discussed below, we limit the number of angles to one per octant.  We also ignore the diffusive flux term as in JSD14; we find it makes little difference.  
In the diffusion limit, the analytic solution is given by (JSD14)
\be
E_r(x,t) = \frac{E_{r,0}}{\sqrt{4 L v_{\rm d} t/\Delta x^2 + 1}}\exp\left(-\frac{{x^2}/{\Delta x^2}}{4 L v_{\rm d} t/\Delta x^2 + 1}\right).\label{eq:diff analytic} 
\ee

In Figure \ref{fig:diff}, we plot $E_r$ for the case of $t = 0.1, 0.5, $ and $1\, L/v_{x,0}$. We overplot the analytic solution given by equation (\ref{eq:diff analytic}).  This Figure shows that our algorithm can calculate the diffusion and advection processes accurately without the splitting of the transport step as done in JSD14.  
}

\subsection{Crossing Beams}\label{sec:crossing}

The propagation of two beams of radiation in vacuum is a straightforward test which can produce qualitatively wrong results depending on the radiative transfer method.  For instance, M1 methods tend to merge two beams of radiation at point that they intersect due to there being a single preferred direction in this moment method (JSD14).  Short characteristics methods \citep{2012ApJS..199....9D,2012ApJS..199...14J} and time dependent angular methods (JSD14), on the other hand, produce correct results.

\begin{figure*}
 \includegraphics[width=0.48\textwidth]{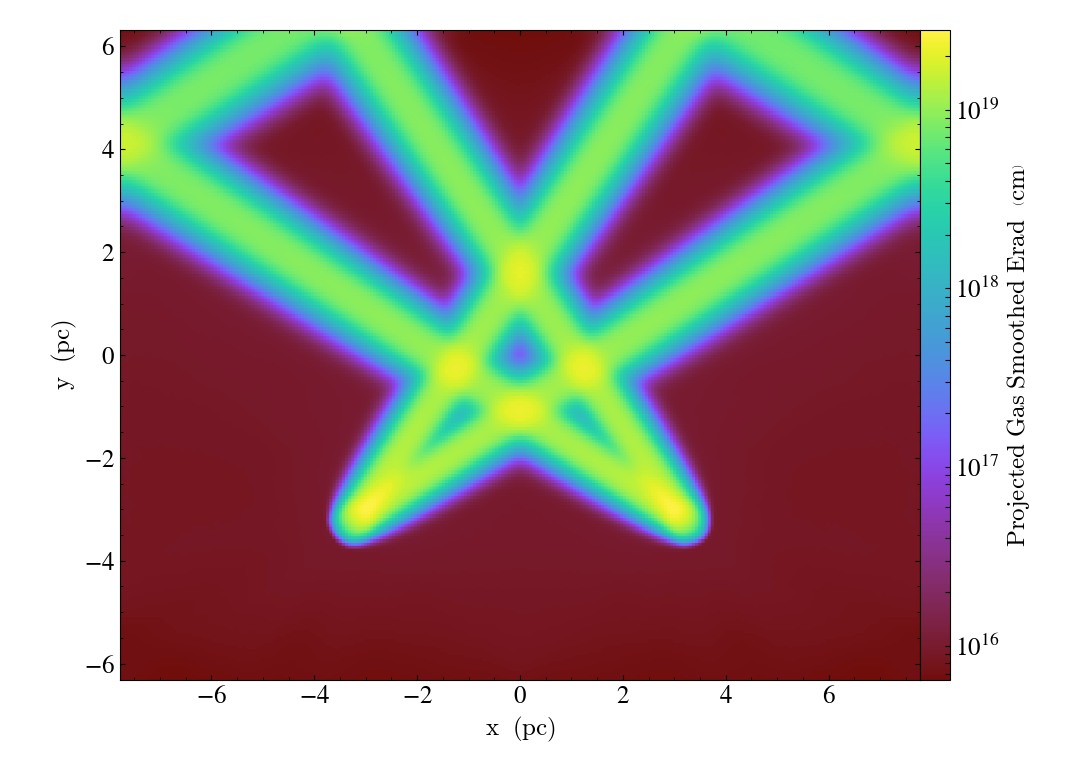}
 \includegraphics[width=0.48\textwidth]{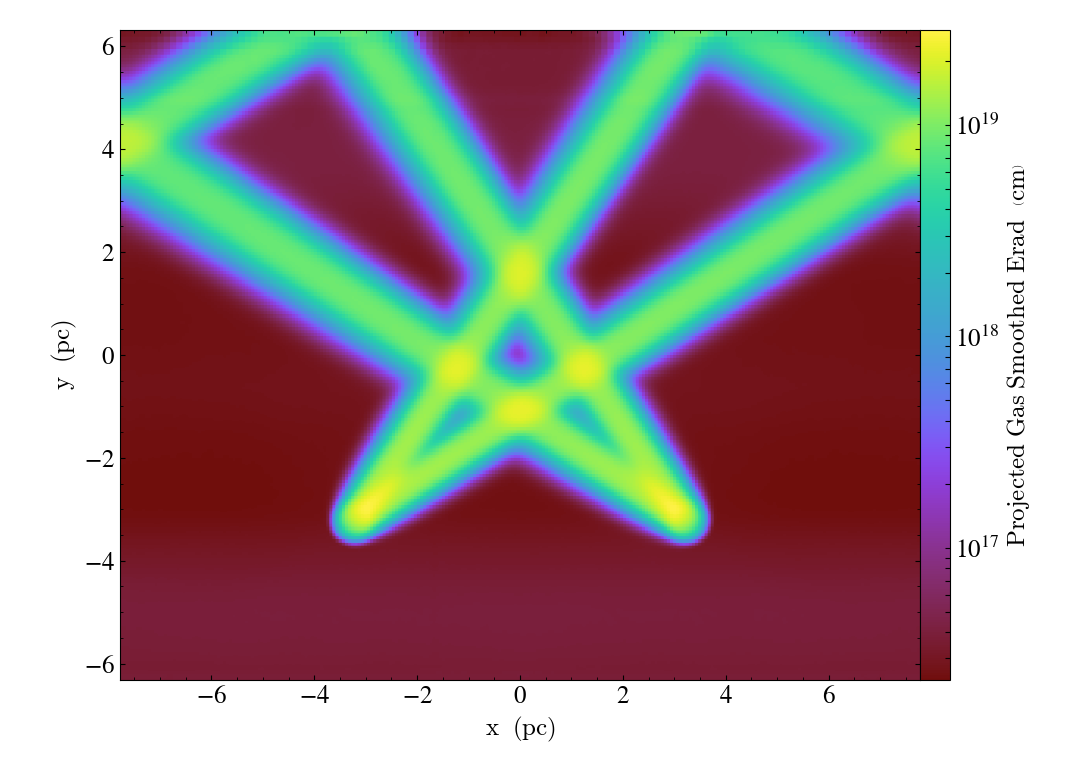}
   \caption{Projected $E_r$ for a stationary fluid and mesh (left) and the moving fluid and mesh (right).  In the right plot, the fluid moves in the +x direction at a velocity of $v=3\,{\rm km\ s^{-1}}$, which is 30\% of the reduced speed of light \rc \label{fig:crossing}}
\end{figure*}

In Figure \ref{fig:crossing}, we show the results of the crossing beam test in our moving-mesh implementation.  Here, we set $\rc = 10\,{\rm km\ s^{-1}}$ in a box that is 16 pc on a side and fill it with $3\times 10^5$ particles by replication a glass distribution of 4,096 particles.  We note that the actual values of the box and reduced speed of light is irrelevant here as the absorption and scattering opacities are minimal, e.g., $\tau < 1$ across the box.  We use 80 angles in the simulation to cover the unit sphere.  In this simulation, we do not perform an implicit solve for the gas temperature as describe in \S~\ref{sec:implicit}, but rather leave it fixed.  This allows us to focus on the propagation of radiation.  We place sources as cylindrical sources along z at $(x,y) = (-10^{19},-10^{19})\,{\rm cm}$ and $(10^{19},-10^{19})\,{\rm cm}$ with directions of $\theta = \pi/4$ and $3\pi/4$ on the left and right cylinder respectively.  We use a radiation temperature of $T_{\rm eff} = 200$ K. Here $\theta$ is defined along the x-y plane.  The opening angle is set such that $\cos\delta\theta < 0.95$, which picks out just two angles per cylindical source.  Finally, we place an absorbing boundary condition at $y=2\times 10^{19}$ cm and periodic boundary conditions in the x-z boundaries.

In the left plot, the fluid has zero velocity and we see that the results are qualitatively correct after several radiation crossing times.  In the right plot, the fluid moves in the +x direction at a velocity of $v=3\,{\rm km\ s^{-1}}$, which is 30\% of the reduced speed of light \rc.  This motion of the fluid moves the cells in a MM code and some effect in light propagation is expected when the velocity of the fluid approaches the reduced speed of light.  Here we find that after beams have propagated there is little difference between the case where mesh generating points are static (left plot) and where they are moving (right plot) at 30\% of the reduced speed of light. Though we do not show it, we did find that the radiation front is distorted in the direction of the mesh generating points' motion. Finally, distortions on the propagation of the beams show up when the mesh generating points approach 50\% of the reduced speed of light or higher, which we also do not show.   However, given the effect of slowly moving meshes (relative to $\rc$) on equilibrium radiation fields is small, the reduced speed of light approximation is excellent when $\rc$ is set to be few times faster than the fastest fluid velocity and/or mesh generating points.

\subsection{Shadow Test}\label{sec:shadow}

Like the crossing beam test, the shadow test has been used to demonstrate the ability of various methods (FLD, M1, RT, VET) to capture the qualitatively correct behavior of photons propagating in multiple directions simultaneously.  This test has been widely use to demonstrate the superiority of M1 methods to capture the shadow cast by one beam compared to FLD, which does not cast shadows \citep{2014MNRAS.441.3177M}.  However, M1 performs more poorly with more than one beam, e.g. multiple shadows. In particular, the closure used in M1 can only account for one direction of propagation.  For more than one beam, the direction is then taken as the intensity weighted direction, which is qualitatively incorrect.  On the other hand, time dependent methods (JSD14) or VET \citep{2012ApJS..199....9D,2012ApJS..199...14J} methods capture the correct qualitative dynamics.

We set up a multiple shadow simulation with a 16 pc side box and populate it with 3.1 M mesh-generating points.  This simulation is larger than the others above because we want to capture the sharp transition between fluids of different densities.  We set an ambient density of $\rho=3\times 10^{-22}\,{\rm g\,cm}^{-3}$ with a mean opacity of $\kapAbsJ=100\,{\rm cm^2\,g}^{-1}$, making it marginally optically thick, $\tau \approx 1$.  The Planck opacity (corresponding to emission) is set much smaller $\kapAbsPl=10^{-2}\,{\rm cm\,g^{-1}}$, so that there is negligible re-emission.  As in the crossing beam test, we adopt periodic boundary conditions along the x and z axis.  We use absorbing boundary conditions for the radiation and periodic boundary conditions for the gas. The exception is the lower y-boundary where we set the boundary to be a fixed radiation input field.  The radiation is set at directions of $\theta = 0.98$ (56 degrees) and $-0.98$, measured from the y-axis, and the radiation temperature is $T_{\rm eff} = 200$ K.
\
At $(x,y) = (0,-3)$ pc, we place a dense elliptical tube that extends in the z-direction.  We set the density of the tube to be 1000x larger than the ambient density so that it is optically thick.  The elliptical tube has a major axis of 4.8 pc and a minor axis of 3 pc, with the major axis oriented along the x-axis.    We adopt the same opening angle of $\cos\delta\theta < 0.98$, which picks out just one angle.  Finally, we place absorbing boundary condition at the top and bottom (in y), but periodic elsewhere.  We ignore the radiation coupling to the hydrodynamics though the hydrodynamics are evolved.

\begin{figure}
 \includegraphics[width=0.5\textwidth]{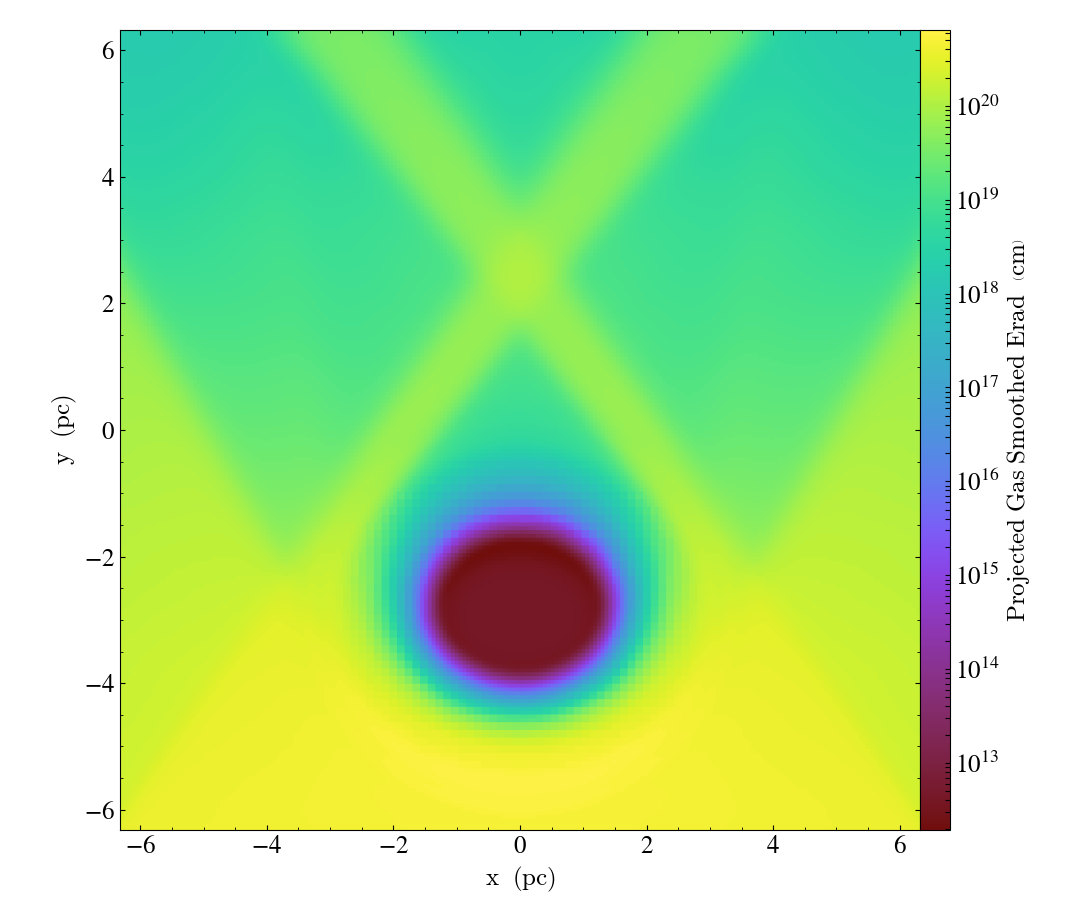}
   \caption{Projected $E_r$ for a multiple shadow test.  Radiation is emitted with a radiation temperature of $T_{\rm eff} = 200\,{\rm K}$ from the bottom boundary with angle $\theta = 0.98$ (56 degrees) and $-0.98$ from the y-axis.  Two shadows will be produced and the shaded beams are also visible. \label{fig:shadow}}
\end{figure}

In Figure \ref{fig:shadow}, we project $E_r$ for the multiple shadow test. The mulitple shadows cast by the tube is qualitative what is expected.  the umbra and penumbra are visible and follow our qualitative expectiations. As argued in JSD14, these results are qualitatively superior to the M1
method, which casts only one shadow.

\subsection{Linear Wave}\label{sec:linear}
The radiative linear wave is a full end-to-end test of the full radiation hydrodynamics algorithm and, thus, we carry out the linear wave test in a 3D domain
based on the dispersion relation \citet{2012ApJS..199...14J} \citep[see also][]{2014ApJS..213....7J,2018ApJ...854..110Z}. As discussed in \citet{2014ApJS..213....7J}, the relevant parameters for this test are $\mathcal{C} = c/c_s$, the ratio between the speed of light and isothermal sound speed, $r = \rc/c$, the ratio between the reduced speed of light and speed of light, $\mathcal{P} = P_{\rm rad}/P_{\rm g}$, the ratio between the radiation and gas pressures, $\tau_{\lambda} = \kappa\rho\lambda$, the optical depth over one wavelength, and $\gamma = 5/3$, the adiabatic gas index.  

The setup in \changaMM assume physical units, so we setup a uniform medium with background density $\rho = 10^{-2}\,{\rm g\,cm}^{-3}$ and box size in the x-direction $L_x = 3\times 10^{10}$ cm.  The box size in the y and z direction are equal, $L_y=L_z$ and set based on the resolution in the x-direction and scaled such that the total number of 3-d mesh generating points remains approximately constant.  These physical units are irrelevant to the problem once everything is scaled to the four parameters described above.  We setup the mesh-generation points with a regular lattice structure, but perturbed randomly by 10\% of the grid spacing to ensure that there are no degeneracies in the Voronoi planes.  We also fix the mesh-generating points in this test as the motion of the mesh generating points (in the regularization step, e.g, see \citealt{Chang+17}) produced noise that can exceed the perturbation of the linear wave. 

Setting the wavelength of the linear wave to be the size of the box in the x-direction, we adopt $\mathcal{C} = 1.45\times 10^3$, $r = 10^{-2}$, and $\mathcal{P} = 1$. In other words, we use a reduced speed of light which is 1\% of the speed of light and a sound speed that is $\approx 0.1$ of the reduced speed of light.  We also set the radiation pressure equal to the gas pressure. We have run tests with different $r$ and $\mathcal{P}$ and the results are consistent with the analytic results (using those parameters) that we describe below.  We use the analytical solutions of \citet{2012ApJS..199....9D} for  $\tau_{\lambda} = 10^{-2}-10^2$ to initialize a linear wave with a dimensionless amplitude of $10^{-3}$.

\begin{figure}
   \includegraphics[width=0.48\textwidth]{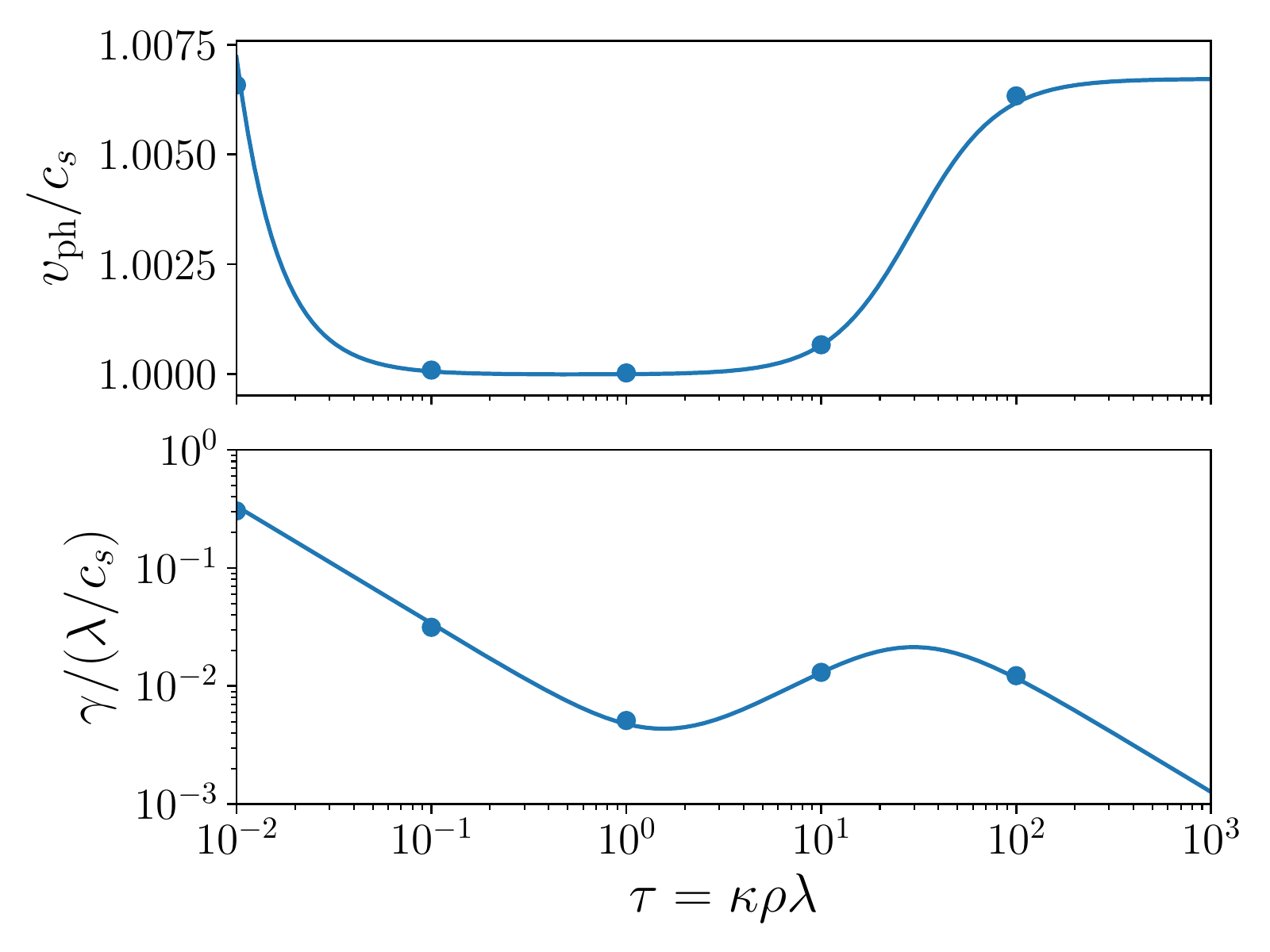}
      \caption{Phase velocity, $v_{\rm ph} = \omega_r/k$, in units of the isothermal sound speed, $c_s$ (top), and damping coefficient, $\gamma$ (bottom), in units of $\lambda/c_s$ for $P_{\rm rad} = P_{\rm g}$, $c/c_s = 1.45\times 10^3$ and $c_r/c = 10^{-2}$ as a function of optical depth across a wavelength, $\tau_w$.  Solid dots are fits from the simulations. \label{fig:dispersion}}
  \end{figure}

In Figure \ref{fig:dispersion}, we compare the measured phase velocity, $v_{\rm ph}$ in units of the isothermal sound speed, $c_s$, and the measured damping rate, $\gamma$, from our \changaMM simulation with an effective grid resolution of 160 in the x-direction with the analytic curves from \citet{2018ApJ...854..110Z}.  We measure the phase velocity and damping rate by evolving our linear wave for one isothermal sound speed crossing time and fitting the linear wave solution for the phase and amplitude.  It is clear from this figure that the analytic results are captured with a high fidelity.  

\begin{figure}
   \includegraphics[width=0.48\textwidth]{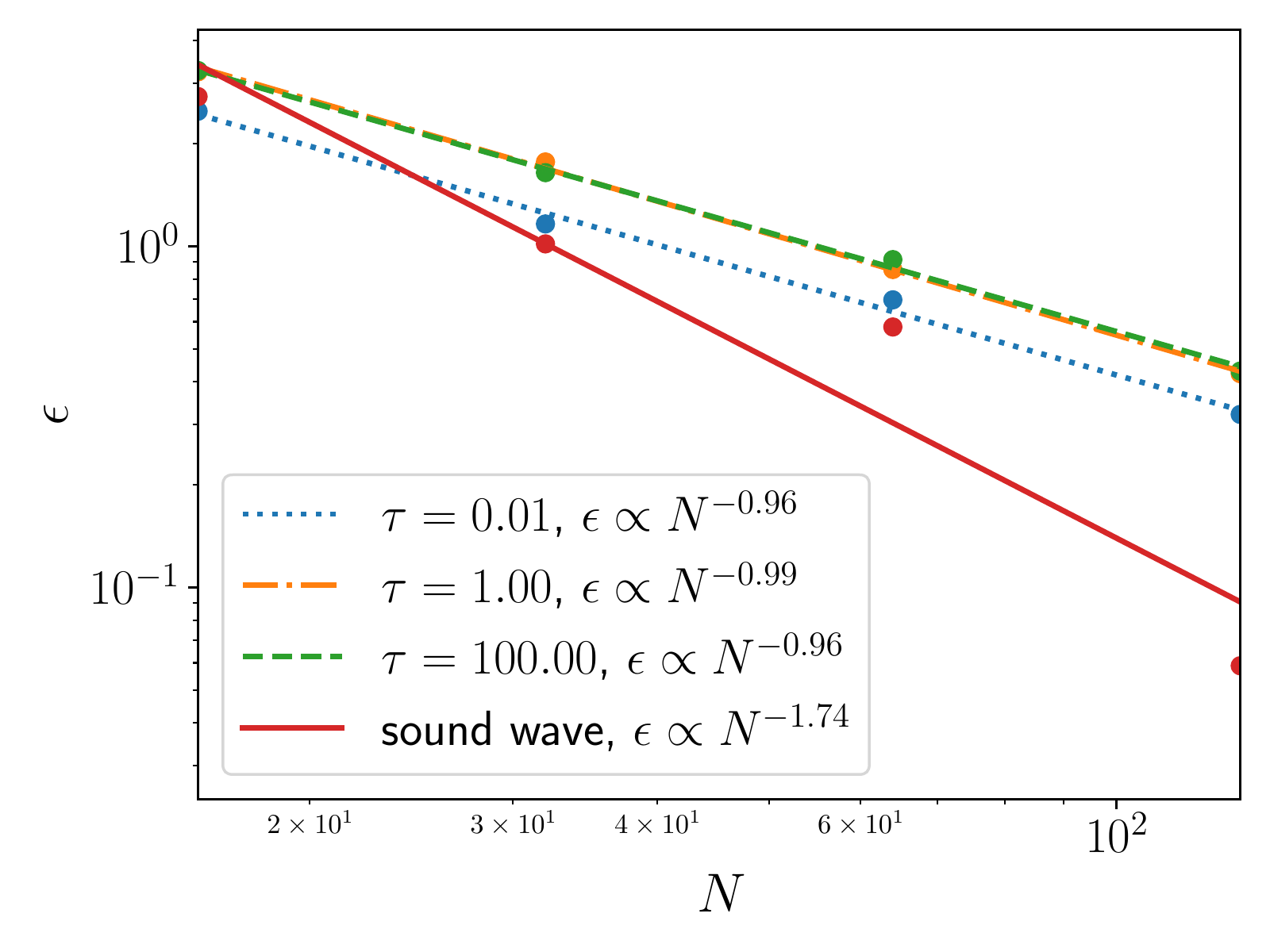}
      \caption{Error (as L1 norm) as a function of x-resolution, $N$, for $\tau = 0.01,\ 1,\ 100$, spanning the range from optically thin to thick.  We also plot the error for a pure acoustic wave, which has an index of $-1.74$, which is close to the expected second order convergence index of $-2$. The convergence of the radiation cases are consistent with first order convergence.\label{fig:convergence}}
  \end{figure}

In Figure \ref{fig:convergence}, we plot the error, $\epsilon$ (with arbitrary normalization) as a function of the number of grid points in the x-direction, e.g., the effective resolution.  Here we define the error, $\epsilon$, as the L1 norm between the analytic and numerical solutions after one isothermal crossing time.  A fit to the error shows that the convergence of the radiative linear wave (in all optical thickness regimes) is approximately a power law with an index of $-1$. We confirm the second order nature of the hydrodynamics solver by showing the L1 norm of a pure acoustic wave have a power law index of -1.74 with respect to linear resolution.  This is nearly the same index found by \citet{2015ApJS..216...35Y}, whose methods of gradient estimation we adopt in \changaMM \citep{Chang+17}. 

Although the integration and reconstruction scheme might be expected to yield second order convergence, the method is only formally first order since the implicit update of the radiation transfer source term is only first order in time.  Hence, our first order convergence is consistent with this expectation.  Nevertheless, we note that JSD14 showed second-order convergence for the case of $\tau=0.01$ and $\tilde{c} = c$, which can be attributed to the limited impact of the radiation source term on the sound wave in this regime.  Our convergence remains only first order for $\tau=0.01$ with either $\tilde{c}=10^{−2} c$ or $c$, motivating further study to see if we can improve convergence in this weakly coupled regime.

\begin{figure}
   \includegraphics[width=0.48\textwidth]{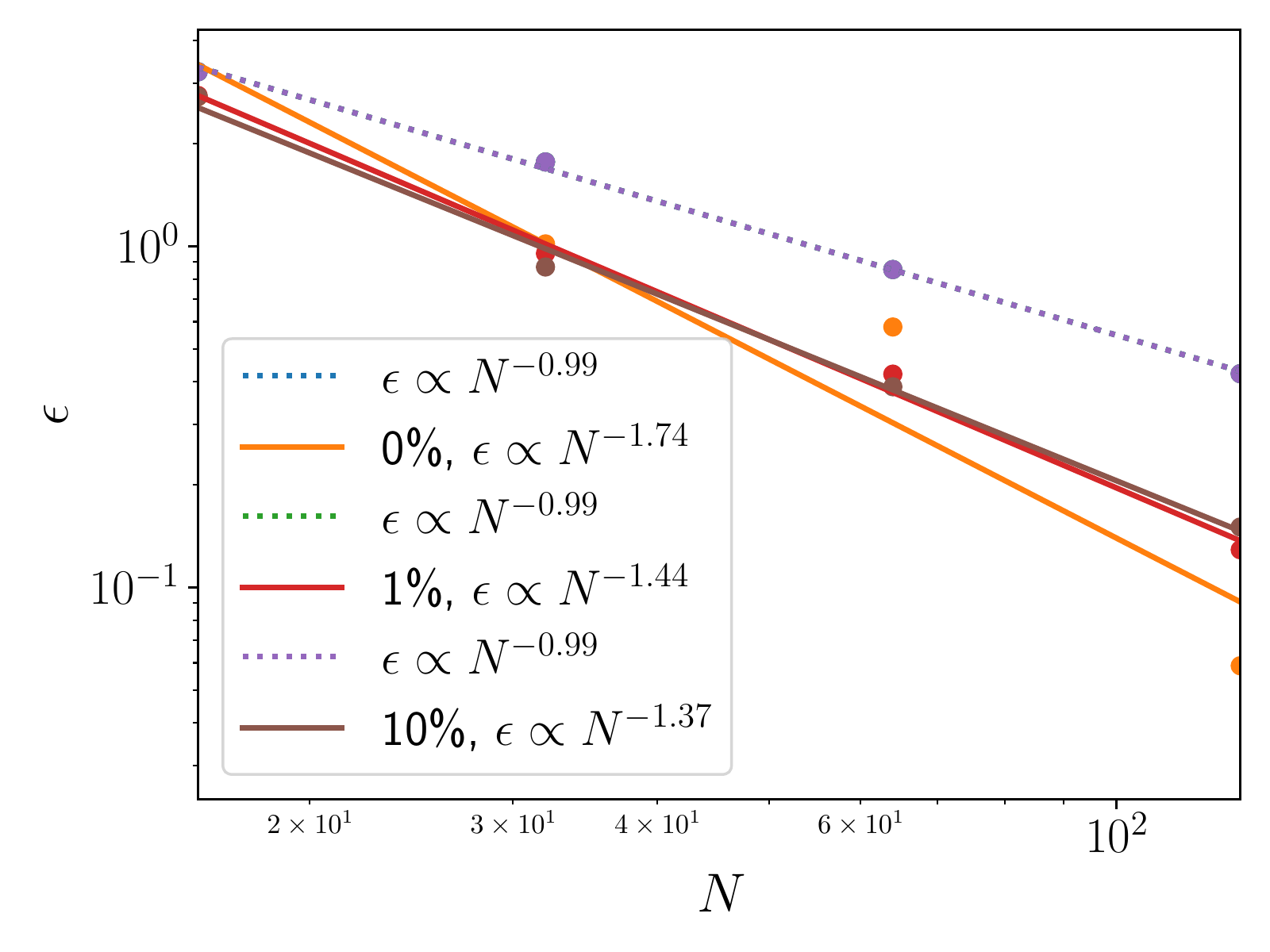}
      \caption{Error (as L1 norm) as a function of x-resolution for $\tau = 1$ and the pure acoustic wave for a nearly cartesian grid $0\%$ deviation and two cartesian grids with a poisson deviation in the mesh generating points of $1\%$ and $10\%$ of the separation, $dx$.  The radiative linear wave is dominated by the first-order convergence of the radiation solver, while the second order convergence of the sound wave is degrade by an increasingly non-regular mesh.\label{fig:convergence-dist}}
\end{figure}

Finally, we explore the effect of mesh regularity on the convergence of our result as previously discussed by \citet{2016MNRAS.455.1134P}.  In Figure \ref{fig:convergence-dist}, we plot the L1 norm for the radiative linear wave with $\tau=1$ and the pure acoustic wave for a nearly cartesian mesh (deviations in the mesh generating points of $10^{-6}$ of the seperation, $dx$) and two cartesian meshes with deviation of 1\% and 10\% of the separation.  For the acoustic wave, we can see that increasing non-regularity of the mesh degrades the second order convergence of the hydrodynamics solver, but the linear wave suffer no such degradation.  Instead, its first-order convergence appears to be dominated by the radiation solver.

\section{Discussion and Conclusions}\label{sec:discussion}

We have adapted JSD14's algorithm for radiation hydrodynamics based on solving the time-dependent RT equation to an unstructured MM and implemented it in the MM code, \changaMM. We solve the specific intensities along different angles and integrate these intensities over angles to find the radiation energy
and momentum source terms that coupled to the hydrodynamic equations. We have tested our implementation on a set of simple problems.  These test problems show that energy and momentum between radiation and matter is conserved by the implicit solver even when the coupling time between radiation and matter is much shorter than the characteristic timestep.  Additionally, the crossing beam and multiple shadow tests demonstrate that the method produces qualitatively correct results in the optically thin limit when multiple sources and obstructions give rise to a complex radiation field.
In particular, these latter two tests demonstrate the importance of the angular distribution of radiation in producing the qualitatively correct dynamics. As stated in JSD14, the principle advantage of this method is that the
angular distribution of photons is calculated self-consistently.  Because we do not use an ad-hoc closure relation as required in FLD
and M1 methods, this method is generally superior in the optically thin case with complex radiation fields. Finally, we demonstrate the fidelity of our radiation hydrodynamics solver in reproducing an analytic results in the linear wave test and converges appropriately with higher resolution.

As mentioned by JSD14, the method of solving the time-dependent RT equations is more straightforward to implement than the method of short characteristics as it does not require a global solve.  In addition JSD14 mentions that this method treats radiation and hydrodynamic variables on a similar footing. Thus, it is straightforward to extend it to curvilinear coordinate systems with nonuniform grids. Here, we recognized their insight by extending it to moving unstructured grids.

Our primary planned applications for this radiation module are dynamical stellar problems such as stellar mergers \citep{2019MNRAS.486.5809P} and tidal disruption events.  Cases where radiation pressure blows material away may require the development of a grid regularization scheme.  Radiation momentum driving tends to push material away from sources, which typically are a small fraction of the volume of a simulation box.  As a result, moving meshes can quickly get distorted without some sort of mesh regularization that either keeps the volume or mass of each cell fairly regular or an algorithm to generate more mesh points to maintain mesh regularity (S10). By contrast gravity tends to pull material from the entire simulation volume into small regions, so the need for such algorithms, while desirable, is not required.  For this reason, we have refrained from simulating radiation feedback problems \citep{2012ApJ...760..155K,2014ApJ...796..107D,2015MNRAS.449.4380R}, which \citet{2019MNRAS.485..117K} has recently demonstrated using AREPO-RT.  We plan to implement a mesh regularization scheme for \changaMM to address this need.

There are a number of improvements that would greatly improve the utility of the radiation solver.  These include the inclusion of magnetic fields, alternative RT schemes, additional mesh generation/regularization improvements, and performance improvements.  Here we note that the recent implementation of RT in Arepo-RT \citep{2019MNRAS.485..117K}, contains an ionization solver.  We have no plans to implement ionization in our current scheme as the relevant science questions that we are interested in does not require it. 

Implementation of alternative RT schemes would permit comparison between this full angular schemes and moment schemes such as FLD or M1.  Here we have mentioned that limitation of M1 in producing the qualitatively correct solutions to various test problems such as the crossing beam test and double shadow test.  At the same time, the full angular implementation we use here suffers from angular discretization which for single sources may be at a disadvantage compared to M1.  A detailed comparison between different RT schemes for specific problems is a problem worthy of further study.

Performance improvements for these various schemes are planned for the near future.  Even in the reduced speed of light approximation, the reduced speed of light is usually at least a factor of a few faster than the fastest velocity, which limits the timestep.  However, the radiative transfer solution can be greatly simplified compared to the hydrodynamic step in that the same Voronoi structure can be assumed to be constant across the step.  As a results, subcycling the radiative transfer step may result in substantial computational savings as \citet{2019MNRAS.485..117K} have recently implemented in AREPO-RT.  

The source code, test problems, and documentation for \changaMM will be available with all the physics (MHD, radiation) in an anticipated future public release of \changa.  We hope that this code will find wide application in astrophysical problems.  We encourage interested parties to contact us if they are interested in using \changaMM before the public release.
%The alert reader will note that we have refrained from a discussion of the performance of our scheme compared to fixed grid Eulerian or SPH methods.  This is intentional and dates back to our original paper introducing \changaMM \citep{Chang+17}.  The main reason is that we have only recently implemented individual timestepping for the MM algorithm for pure hydrodynamics (Chang \& Prust, in prep), wherea the SPH solver has always had individual timestepping.  For problems with a large dynamic range, individual timesteps are one of the biggest algorithmic improvement in the performance of these codes.  In the near future, we will adapt this radiation hydrodynamics algorithm for individual timestepping.

\section*{Acknowledgements}

We thank Norman Murray and Paul Duffell for useful discussions. 
We thank Jim Stone for useful discussions and for permission to use various code segments of Athena and Athena++ in \changaMM. We thank the anonymous reviewer for a constructive report. 
PC is supported by the NASA ATP program through NASA grant NNH17ZDA001N-ATP, NSF CAREER grant AST-1255469, and the Simons Foundation.
SWD acknowledge support from NSF grant AST-1616171 and an Alfred P. Sloan Research Fellowship.
YFJ was supported in part by the National Science Foundation under Grant No. NSF PHY-1748958.
We also used the Extreme Science and Engineering Discovery Environment (XSEDE), which is supported by National Science Foundation (NSF) grant No. ACI-1053575. We also acknowledge the Texas Advanced Computing Center (TACC) at The University of Texas at Austin for providing HPC resources that have contributed to the research results reported within this paper. URL: \url{http://www.tacc.utexas.edu}. We also use the yt software platform for the analysis of the data and generation of plots in this work \citep{yt}.The Flatiron Institute is supported by the Simons Foundation.

\bibliographystyle{mnras}
\bibliography{references}

%%%%%%%%%%%%%%%%%%%%%%%%%%%%%%%%%%%%%%%%%%%%%%%%%%

%%%%%%%%%%%%%%%%% APPENDICES %%%%%%%%%%%%%%%%%%%%%

% \appendix

% \section{Radiatively Damped Acoustic Waves}
%
% In this section we derived the linear theory of radiatively damped.  While the derivation is straightforward and similar to that of a sound wave, we include it in the Appendix for clarity of exposition.
%
% We begin first with the 1-d Euler equations written in primitive form and ignoring gravity:
% \be
% \ddt{\rho} + \ddx{\rho v} = 0,\\
% \ddt{v} + v\ddx{v} = -\frac 1 {\rho}\ddx{P},\\
% \ddt{\rho\epsilon} + v\ddx{\rho\epsilon} + (P + \rho\epsilon)\ddx{v} = -c\rho\kappa\left(aT^4 - E_{\rm rad}\right),
% \ee
% We also include the radiation term.
% \be
% \ddt{E_{\rm rad}} = c\rho\kappa\left(aT^4 - E_{\rm rad}\right)
% \ee
% With the expansion around $v=0$, $\rho$, we have
% \be
% -i\omega\delta \rho + \rho_0 ik\delta v = 0,\\
% -i\omega\delta v  = -\frac {ik\delta P} {\rho},\\
% -i\omega\delta P + \gamma P ik\delta v = -(\gamma - 1)c\rho\kappa\left(E_{\rm rad}4\frac{\delta T}{T} - \delta E_{\rm rad}\right),\\
% -i\omega \delta E_{\rm rad} = c\rho\kappa\left(E_{\rm rad}4\frac{\delta T}{T} - \delta E_{\rm rad}\right)
% \ee
% where we have used ${\rho\epsilon} = P/(\gamma- 1)$.  We adopt an ideal gas equation of state, e.g., $\delta P/P = \delta \rho/\rho + \delta T/T$

%%%%%%%%%%%%%%%%%%%%%%%%%%%%%%%%%%%%%%%%%%%%%%%%%%

% Don't change these lines
\bsp	% typesetting comment
\label{lastpage}
\end{CJK*}
\end{document}